\documentclass[pre,floatfix,longbibliography,superscriptaddress,onecolumn,final,notitlepage]{revtex4-1}

\usepackage[utf8]{inputenc}
\usepackage[T1]{fontenc}
\usepackage{multirow}
\usepackage{subcaption}
\usepackage{tabularx}
\usepackage{graphicx,caption}
\usepackage{amsfonts,amsmath,amssymb}
\usepackage{comment}
\usepackage{lineno}
\usepackage{hyperref}
\usepackage[explicit]{titlesec}

\newcommand{\be}{\begin{equation}}
\newcommand{\ee}{\end{equation}} 
\newcommand{\bea}{\begin{eqnarray}}
\newcommand{\eea}{\end{eqnarray}}

\renewcommand{\v}[1]{\mathbf{#1}} 

\newcommand{\f}[2]{\frac{#1}{#2}}

\newcommand{\ccup}[1]{\left\{#1\right\}}

\newcommand{\bup}[1]{\left(#1\right)}
\newcommand{\rup}[1]{\left[#1\right]}

\renewcommand{\t}[1]{\tilde{#1}}

\newcommand{\id}{\mathbb{I}}

\renewcommand{\ref}[1]{[\ref{#1}]}

\usepackage{tabularx}

\usepackage{amsmath,fixmath}
\usepackage{comment}
\usepackage{lineno}
\usepackage{hyperref}
\usepackage[explicit]{titlesec}

\usepackage{graphicx}
\usepackage[labelfont=bf]{caption}
\usepackage[format=hang]{subcaption}

\usepackage{algorithm,algorithmic}

\hypersetup{
 bookmarks=true, 
 unicode=false, 
 pdftoolbar=true, 
 pdfmenubar=true, 
 pdffitwindow=false, 
 pdfstartview={FitH}, 
 pdftitle={}, 
 pdfauthor={}, 
 pdfcreator={}, 
 pdfproducer={}, 
 pdfkeywords={} {} {}, 
 pdfnewwindow=true, 
 colorlinks=true, 
 linkcolor=red, 
 citecolor=blue, 
 filecolor=magenta, 
 urlcolor=cyan 
}
\usepackage[algoruled,algo2e]{algorithm2e}
\usepackage{sidecap}
\usepackage{boldline}
\usepackage{setspace}

\usepackage[normalem]{ulem} 

\newcommand{\mt}{\mbox{{\small MULTITENSOR}}}
\newcommand{\mtcov}{\mbox{{\small MTCOV}}}
\newcommand{\cesna}{\mbox{{\small CESNA}}}
\newcommand{\nc}{\mbox{{\small NC}}}

\begin{document}

\title{Community detection with node attributes in multilayer networks}

\author{Martina Contisciani}
\email{martina.contisciani@tuebingen.mpg.de}
\affiliation{Max-Planck Institute for Intelligent Systems, Cyber Valley, Tuebingen 72076, Germany}

\author{Eleanor A. Power}
\email{e.a.power@lse.ac.uk}
\affiliation{London School of Economics and Political Science, Department of Methodology, London, WC2A 2AE, United Kingdom}

\author{Caterina De Bacco}
\email{caterina.debacco@tuebingen.mpg.de}
\affiliation{Max-Planck Institute for Intelligent Systems, Cyber Valley, Tuebingen 72076, Germany}

\begin{abstract}
\textsl{Community detection in networks is commonly performed using information about interactions between nodes. Recent advances have been made to incorporate multiple types of interactions, thus generalising standard methods to multilayer networks. Often, though, one can access additional information regarding individual nodes, attributes, or covariates. A relevant question is thus how to properly incorporate this extra information in such frameworks.  Here we develop a method that incorporates both the topology of interactions and node attributes to extract communities in multilayer networks. We propose a principled probabilistic method that does not assume any \textit{a priori} correlation structure between attributes and communities but rather infers this from data. This leads to an efficient algorithmic implementation that exploits the sparsity of the dataset and can be used to perform several inference tasks; we provide an open-source implementation of the code online. We demonstrate our method on both synthetic and real-world data and compare performance with methods that do not use any attribute information. We find that including node information helps in predicting missing links or attributes. It also leads to more interpretable community structures and allows the quantification of the impact of the node attributes given in input. }
\end{abstract}

\maketitle

\thispagestyle{empty}


\section*{Introduction}
\label{sec: Introduction}

Community detection is a fundamental task when investigating network data. Its goal is to cluster nodes into communities and thus find large-scale patterns hidden behind interactions between many individual elements. 

The range of application of this problem spans several disciplines. For instance, community detection has been used in sociology  to analyze terrorist groups in online social networks \cite{waskiewicz2012friend}; in finance to detect fraud events in telecommunication networks \cite{pinheiro2012community}; in engineering to refactor software packages in complex software networks \cite{pan2013refactoring}; and in biology to investigate lung cancer \cite{bechtel2005lung} and to explore epidemic spreading processes \cite{chen2012epidemic}. 
In recent years, the variety of fields interested in this topic has broadened and the availability of rich datasets is increasing accordingly. However, most research approaches use only the information about interactions among nodes, in other words the network topology structure. This information can be complex and rich, as is the case for multilayer networks where one observes different types of interactions. For instance, in social networks, interactions could entail exchanging goods, socializing, giving advice, or requesting assistance. Most network datasets, however, contain additional information about individuals, attributes which describe their features, for instance their religion, age, or ethnicity. Node attributes are often neglected \textit{a priori} by state-of-the-art community detection methods, in particular for multilayer networks. They are instead commonly used \textit{a posteriori}, acting as candidates for ``ground-truth'' for real-world networks to measure the quality of the inferred partition \cite{traud2011comparing,newman2006modularity}, a practice that can also lead to incorrect scientific conclusions \cite{Peele1602548}.
It is thus a fundamental question how to incorporate node attributes into community detection in a principled way. This is a challenging task because one has to combine two types of information \cite{yang2013community}, while evaluating the extent to which topological and attribute information contribute to the network's partition \cite{falih2018community}. 

To tackle these questions, we develop \mtcov, a mathematically rigorous and flexible model to address this problem for the general case of multilayer networks, i.e., in the presence of different types of interactions. The novelty of this model relies on a principled combination of the multilayer structure together with node information to perform community detection. To the best of our knowledge, \mtcov\text{} is the first overlapping community detection method proposed for multilayer networks with node attributes. 
The model leverages two sources of information, the topological network structure and node covariates (or attributes), to partition nodes into communities. It is flexible as it can be applied to a variety of network datasets, whether directed, weighted, or multilayer and it outputs overlapping communities, i.e., nodes can belong to multiple groups simultaneously. 
  In addition, the model does not assume any \textit{a priori} correlation structure between the attributes and the communities. On the contrary, the contribution of the attribute information is quantitatively given as an output of the algorithm by fitting the observed data. The magnitude of this contribution can vary based on the dataset. Even if this is not very high (for instance if the attributes are noisy or sparse) the model is nevertheless able to use this extra information to improve performance. At the same time, if incorporating attribute information hurts inference tasks, the model will downweigh this contribution and instead use mostly the topological network structure. 

Our method allows domain experts to investigate particular attributes and select relevant community partitions based on what type of node information they are interested in studying. In fact, by choosing the input data, we can drive the algorithm to select for communities that are more relevant to the attribute under study. If the attribute hurts performance and is consequently downweighted by the algorithm, this can be used as a signal that the attribute might not correlate well with any partition, given the remaining topological information available, and thus inform the expert accordingly.

We study \mtcov\text{} on synthetic multilayer networks, a variety of single-layer node-attributed real networks and several real multilayer networks of social support interactions in two Indian villages. We measure performance based on prediction tasks and overlap with ground-truth (when this is known). For single-layer networks, we compare the performance of \mtcov\text{} to state-of-the-art community detection algorithms with node attributes; for multilayer networks, we test against a state-of-the-art algorithm that does not use any node attribute information and measure the extent to which knowing both types of information helps inference. We find that \mtcov\text{} performs well in predicting missing links and attributes. It also leads to more interpretable community structures and allows the quantification of the impact of the node attributes given as input.

To summarize, we present \mtcov\text{}, a new method that incorporates both the topology of interactions and node attributes to extract communities in multilayer networks. It is flexible, efficient and it has the property of quantitatively estimating the contributions of the two sources of information. It helps domain experts to investigate particular attributes and to better interpret the resulting communities. Moreover, by including relevant node attributes, it boosts performance in terms of edge prediction.

\textbf{Related Work.} Several methods have been proposed to study community detection in networks \cite{fortunato2010community}. In particular, we are interested in those valid for multilayer networks \cite{de2013}. These generalize single-layer networks in that they can model different types of interactions and thus incorporate extra information that is increasingly available. Among these, we focus on generative models for multilayer networks \cite{Multitensor,schein2015,schein2016bayesian,valles2016,stanley2015,peixoto2015,paul2015}, which are based on probabilistic modeling like Bayesian inference or maximum likelihood optimization.  These are flexible and powerful in terms of allowing multiple inference tasks, injecting domain knowledge into the theoretical framework, and being computationally efficient.
However, the majority of these methods do not consider node attributes as input along with the network information. In fact, the few methods developed for community detection in multilayer networks with node attributes are based on first aggregating the multilayer network into a single layer, either by combining directly the adjacency matrices of each layer \cite{gheche2018orthonet} or by using similarity matrices derived from them along with the node attributes \cite{papadopoulos2015, papadopoulos2017}. In the context of data mining, a similar problem can be framed for learning low dimensional representations of heterogeneous data with both content and linkage structure (what we call attributes and edges). This is tackled by using embeddings extracted via deep architectures \cite{chang2015}, which is rather different than our approach based on statistical inference. Our problem bears some common ground with the one studied by Sachan et al. \cite{TUCM} for extracting communities in online social networks, where users gather based on common interests; they adopt a Bayesian approach, but with a rather different goal of associating different types of edges to topics of interest.
A related but different problem is that of performing community detection with node attributes on multiple independent networks \cite{sweet2018estimating, Signorelli_2019}; this differs with modeling a single multilayer network in that it assumes that covariates influence in the same way all the nodes in a network but in a different way the various networks in the ensemble.
For single-layer networks, there has been more extensive work recently on incorporating extra information on nodes \cite{newman2016structure,yang2013community,bothorel2015clustering,zhang2016community,hric2016network,stanley2019stochastic,emmons2019map,sweet2018estimating, xu2012model, bu2017dynamic}. 
Among those adopting probabilistic modeling, some incorporate covariate information into the prior information of the latent membership parameters \cite{sweet2018estimating, tallberg2004bayesian,white2016mixed}, while others include covariates in an additive way along with the latent parameters \cite{ airoldi2011confidence,sweet2015incorporating}, so that covariates influences the probability of interactions independently of the latent membership. 

These works show the impact of adding nodes attributes in community detection \textit{a priori} into the models to uncover meaningful patterns. One might then be tempted to adopt such methods also in multilayer networks by collapsing the topological structure into a suitable single network that can then be given in input to these single-layer and node-attributed methods as done by Gheche et al. \cite{gheche2018orthonet}. However, collapsing a multilayer network often leads to important loss of information, and one needs to be careful in determining when this collapse is appropriate and how it should be implemented, as shown for community detection methods without attribute information \cite{taylor2016enhanced,taylor2017super}.
Thus the need of a method that not only incorporates different types of edges but also node attributes.


\section*{Results}
\label{sec: Results}

We test \mtcov's ability to detect communities in multilayer networks with node attributes by considering both synthetic and real-world datasets. We compare against \mt\text{} \cite{Multitensor}, an algorithm similar to ours but that does not include node attributes.
We also test \mtcov's performance on single-layer networks, as the mathematical framework behind \mtcov\text{} still applies. Given this potential use and the paucity of algorithms suitable for comparison for multilayer networks, such comparisons assess the general utility of \mtcov\text{}.

\subsection*{Multilayer synthetic networks with ground-truth} 
\begin{figure}[htp]
  \centering
    \includegraphics[width=1\linewidth]{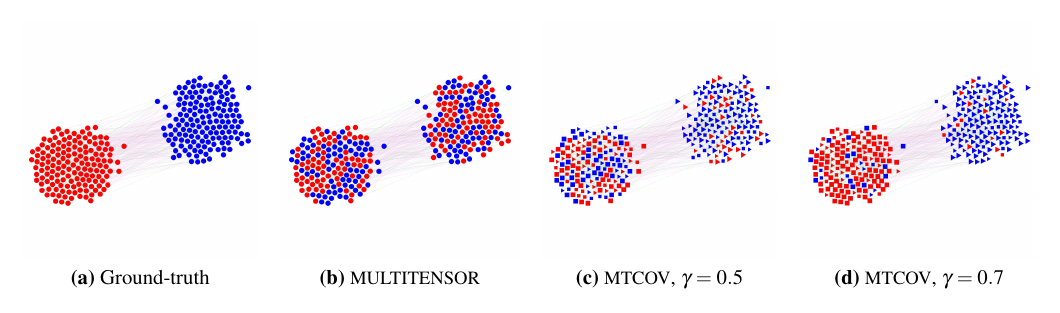}
\caption{Partition of a synthetic multilayer network with attributes. We generated synthetic directed multilayer networks using a stochastic block model, that aligns with $G_1$. To illustrate, here we do the equivalent task on a smaller network of size N = 299, C = 2 communities of equal-size unmixed group membership and L = 2 layers, of which one assortative (green) and one disassortative (pink); (a) the ground-truth partition; (b-d) the communities inferred by three different methods:  (b) \mt, an algorithm without attributes, (c) \mtcov\text{} using the network structure and the attributes with the same proportion, i. e. $\gamma=0.5$ and (d) \mtcov\text{}  using mostly the attribute structure, i.e. $\gamma=0.7$. Colors denote the inferred partition; the attributes in (c) and (d) are generated by matching them with true community assignments for the $50\%$ and $70\%$ of the nodes respectively, and chosen uniformly at random from the non-matching values; square and triangle denote the synthetic dummy attribute  (squares are matched with the red group, triangles with the blue) and the size of the node shows the nodes matched with the true community (bigger means deterministic match, smaller means uniform at random match). We use the matrix $U$ for the membership. 
}     
\label{fig:partition1}
\end{figure}

To illustrate the flexibility and the robustness of our method, we generate multilayer synthetic networks with different kinds of structures in the various layers adapting the protocol described in De Bacco et al. \cite{Multitensor} to accommodate node attributes. We generate attributes as done in Newman and Clauset \cite{newman2016structure}: we match them with planted communities in increasing ratios varying from 0.3 to 0.9; these values correspond also to the $\gamma$ parameters that we fix for \mtcov.
Specifically, we generate three types of directed networks using a stochastic block model \cite{holland1983stochastic}, all with $C=2$ communities of equal-size unmixed group membership and $N=1000$ nodes, but with different numbers and kinds of layers, similar to De Bacco et al. \cite{Multitensor}. The first network ($G_{1}$) has $L=2$ layers, one assortative ($W^{\alpha}$ has higher diagonal entries) and one disassortative ($W^{\alpha}$ has higher off-diagonal entries); the second ($G_{2}$) has $L=4$ layers, two assortative and two disassortative and the third ($G_{3}$) has $L=4$ layers, one assortative, one disassortative, one core-periphery ($W^{\alpha}$ has higher diagonal entries but one of the two is bigger than the other) and one with biased directed structure ($W^{\alpha}$ has higher off-diagonal entries but one of the two is bigger than the other). 
 We generate ten independent samples of each of these types of networks and use all the evaluation metrics described in the Methods section in the presence of ground-truth. We use the membership inferred by the algorithms using the best maximum likelihood fixed point over 10 runs with different random initial conditions. As shown in Table \ref{tab: multisynthetic}, \mtcov\text{} performs significantly better than \mt\text{} on the first and second network. This suggests that incorporating attribute information can significantly boost inference, with an increasing benefit for a smaller number of layers. Figure \ref{fig:partition1} shows an example of this result. Notice that $G_{2}$ requires a smaller match ($\gamma=0.5)$ between attributes and communities than $G_{1}$ ($\gamma=0.7)$ to achieve similar performance. $G_{1}$ and $G_{2}$ have similar structure but the second has twice as many layers. Thus, increasing the number of layers may require less contribution from the extra information of the attributes, a possible advantage for multilayer networks. This intuition is reinforced by noticing not only that the best performance is achieved for $\gamma<0.9$, but also that both the algorithms perform very well in the third network, regardless of the value of the match between attributes and communities. Contrary to $G_{2}$, $G_{3}$ has a different structure in each of the 4 layers. This diversity can be even more beneficial than having more but correlated layers (as in $G_{1}$ \textit{vs} $G_{2}$).  
 These synthetic tests demonstrate the impact of leveraging both node attributes and topological information: when topological structure is not very informative (as in $G_{1}$ with only two layers), adding node attributes can significantly help in recovering the communities. In contrast, when topological information is more complex (as in $G_{3}$ where all layers are different), properly combining the different layers' structures can compensate for a limited access to extra information on nodes. Overall, this shows the need for methods suited for exploiting various sources of information and the complexity behind multilayer networks.

\begin{table}[h!]
\centering
\renewcommand{\arraystretch}{1.2}
\resizebox{\textwidth}{!}
{\begin{tabular}{l|cccc|cccc|cccc}
\clineB{1-13}{3}
\multicolumn{1}{c}{\textbf{}}        & \multicolumn{4}{c|}{\textbf{$G_{1}$}}                                        & \multicolumn{4}{c|}{\textbf{$G_{2}$}}  & \multicolumn{4}{c}{\textbf{$G_{3}$}}         \\
\multicolumn{1}{l|}{\textbf{Method}} & \textbf{F1-score}           & \textbf{Jaccard}           & \textbf{CS} & \textbf{L$_1$}       & \textbf{F1-score}           & \textbf{Jaccard}           & \textbf{CS} & \textbf{L$_1$}  & \textbf{F1-score}           & \textbf{Jaccard}           & \textbf{CS} & \textbf{L$_1$}  \\ \hline
\multicolumn{1}{l|}{\mt}   & 0.512$\pm$0.006  &0.344$\pm$0.006  &0.585$\pm$0.005 &0.492$\pm$0.004 
& 0.514$\pm$0.006&0.346$\pm$0.06 &0.614$\pm$0.005 &  0.490$\pm$0.005
&\textbf{0.999$\pm$0.001}&\textbf{0.998$\pm$0.001}&\textbf{0.991$\pm$0.001}&0.063$\pm$0.002\\
\multicolumn{1}{l|}{\mtcov\_0.3}   & 0.7$\pm$0.2& 0.5$\pm$0.2 &0.7$\pm$0.1 &0.4$\pm$0.1   
& 0.8$\pm$0.2&0.7$\pm$0.2 &0.8$\pm$0.1 &0.3$\pm$0.2  
&0.995$\pm$0.002&0.990$\pm$0.004&0.984$\pm$0.002&0.080$\pm$0.004\\
\multicolumn{1}{l|}{\mtcov\_0.5}   &0.6$\pm$0.1&0.5$\pm$0.2  & 0.7$\pm$0.1& 0.4$\pm$0.1 
& 0.992$\pm$0.005&0.985$\pm$0.009 &0.986$\pm$0.004 &0.064$\pm$0.004  
&0.996$\pm$0.002&0.992$\pm$0.004&0.985$\pm$0.002&0.079$\pm$0.004\\
\multicolumn{1}{l|}{\mtcov\_0.7}   & \textbf{0.988$\pm$0.002}& \textbf{0.976$\pm$0.004} &\textbf{0.977$\pm$0.002} & 0.079$\pm$0.003  
& \textbf{1.$\pm$0.}&\textbf{1.000$\pm$0.001} &\textbf{0.991$\pm$0.001} &0.062$\pm$0.002 
&0.994$\pm$0.002&0.988$\pm$0.004&0.982$\pm$0.001&0.087$\pm$0.002\\
\multicolumn{1}{l|}{\mtcov\_0.9}   & 0.958$\pm$0.003&0.920$\pm$0.005  & \textbf{0.977$\pm$0.001}& \textbf{0.050$\pm$0.002}  
& 0.992$\pm$0.002&0.984$\pm$0.004 &0.988$\pm$0.001 &\textbf{0.050$\pm$0.002}  
&0.976$\pm$0.003&0.952$\pm$0.006&0.982$\pm$0.002&\textbf{0.051$\pm$0.003}\\
\clineB{1-13}{3}
\end{tabular}}
\caption{Performance of algorithms \mt\text{} and \mtcov\text{} on synthetic multilayer networks with attributes. We use  different matches (one per row, e.g. $\mtcov\_{0.3}$ denotes a match of 0.3, this is also the value we use to fix $\gamma$) between attributes and planted communities on synthetic directed multilayer networks. Results are averages and standard deviations over 10 networks samples for each network type $G_{m}$, $m=1,2,3$; we take the average performance over the incoming and outgoing memberships, i.e., the matrices $U$ and $V$, and the best performances are in boldface. Networks are generated with stochastic block model with $C=2$, $N=1000$ and average degree $k=4$. 
Denote $W^{a}$, $W^{d}$, $W^{cp}$ and $W^{bd}$, the affinity matrices of the assortative, disassortative, core-periphery and the biased directed layers respectively. Then, their entries are $w^{a}_{11}=w^{a}_{22}=w^{d}_{12}=w^{d}_{21}=w^{cp}_{11}=w^{bd}_{12}=\f{kC}{N}$, $w^{a}_{12}=w^{a}_{21}=w^{d}_{11}=w^{d}_{22}=w^{cp}_{12}=w^{cp}_{21}=w^{bd}_{11}=w^{bd}_{22}=0.1\times\f{kC}{N}$ and $w^{cp}_{22}=w^{bd}_{12}=0.03\times \f{kC}{N}$. The F1-score, Jaccard, CS and L$_1$ are performance metrics as defined in the Methods section.}
\label{tab: multisynthetic}
\end{table}

\subsection*{Multilayer social support network of rural Indian villages}
We demonstrate our model beyond synthetic data by applying it to social support networks of two villages in Tamil Nadu, India, which we call by the pseudonyms ``Te\underline{n}pa\d{t}\d{t}i'' (Ten) and ``A\underline{l}ak\={a}puram'' (Ala) \cite{power2015, power2017, power2019}. Data were collected in the form of surveys where adult residents were asked to nominate those individuals who provided them with various types of support, including running errands, giving advice, and lending cash or other household items. These were collected in two rounds, one in 2013 and the other in 2017. Each type of support corresponds to a layer in the network; we consider only those layers present in both rounds, for a total of $L=6$ layers. After pre-processing the data, by considering only those individuals who had at least one outgoing edge and removing self-loops, the resulting networks have the size reported in Table \ref{tab: Statistics}. In addition, several attributes were collected, which include information about age, religion, caste, and education level. 
Ethnographic observation in these villages \cite{power2015} and previous analyses \cite{power2017, power2019} suggest that social relations are strongly structured by religious and caste identity, with these divisions shaping where people live, who they marry, and who they choose to associate with. 
In other words, they suggest a dependence between the attributes Religion and Caste and the mechanisms driving edge formation in these social support networks. Motivated by these insights, here we consider the attributes Caste and Religion and add them into the model. In addition, we test the importance of variables that we expect to be less informative, such as Gender and Age. The latter, being continuous, is also an example of a non-categorical variable. Provided it has a finite range, as it is the case for Age, we can encode it into categorical by binning its values. Here we use equal bins of size 5 years.

\begin{table}[ht]
\centering
\renewcommand{\arraystretch}{1.2}
\begin{tabular}{ccccccccc}
\clineB{1-9}{3}
\textbf{Village} & \textbf{Year} & \textbf{Nodes} &$ \mathbf{ Edges}$ & $\mathbf{\langle k \rangle}$& \textbf{Caste} & \textbf{Religion}& \textbf{Age}& \textbf{Gender}\\ \hline
A\underline{l}ak\={a}puram                & 2013          & 419                & 4161& 20 & 14 &        3      &11&    2     \\
              & 2017          & 441             & 5578            & 25        & 13 &   3   &12&    2 \\
Te\underline{n}pa\d{t}\d{t}i               & 2013          & 362                &3374               & 19  & 11 &    2     &11    &    2      \\
             & 2017          & 346                 & 3806              & 22          & 9 &2     &12 &    2  \\
\clineB{1-9}{3}
\end{tabular}
\caption{Network summary statistics for the four social support networks of Indian villages. Each has the same set of 6 layers and Edges are the total over them; $\langle k \rangle$ is the average degree per node on the whole multilayer network. The columns Caste, Religion, Age and Gender are the number of different categories observed in each network for their respective attribute.
}
\label{tab: Statistics}
\end{table}

Without assuming \textit{a priori} any ground-truth, we measure performance using the AUC and accuracy as explained in the Methods section. 
We compare with \mt\text{}  to measure the extent to which adding the attributes helps predicting edges and attributes; in addition, in terms of accuracy values, we consider two baselines for further comparisons: i) a uniform at random probability over the number of possible categories (RP); and ii) the maximum relative frequency of the attribute value appearing more often (MRF). We fix hyperparameters using 5-fold cross-validation along with grid-search procedure (see ``Cross-validation tests and hyperparameter settings'' subsection for more details). We obtain values of $\gamma \in [0.2, 0.9]$, signalling relevant correlations between attributes and communities. For details, see Supplementary Table S2. Empirically, we observe that when $\gamma>0.5$ the algorithm achieves better performance in terms of link and attribute prediction by well balancing the log-likelihood of the attribute dimension and the one of the network structure. 

\begin{figure}[phbt]
  \begin{minipage}[c]{0.45\linewidth}
    \includegraphics[width=\linewidth]{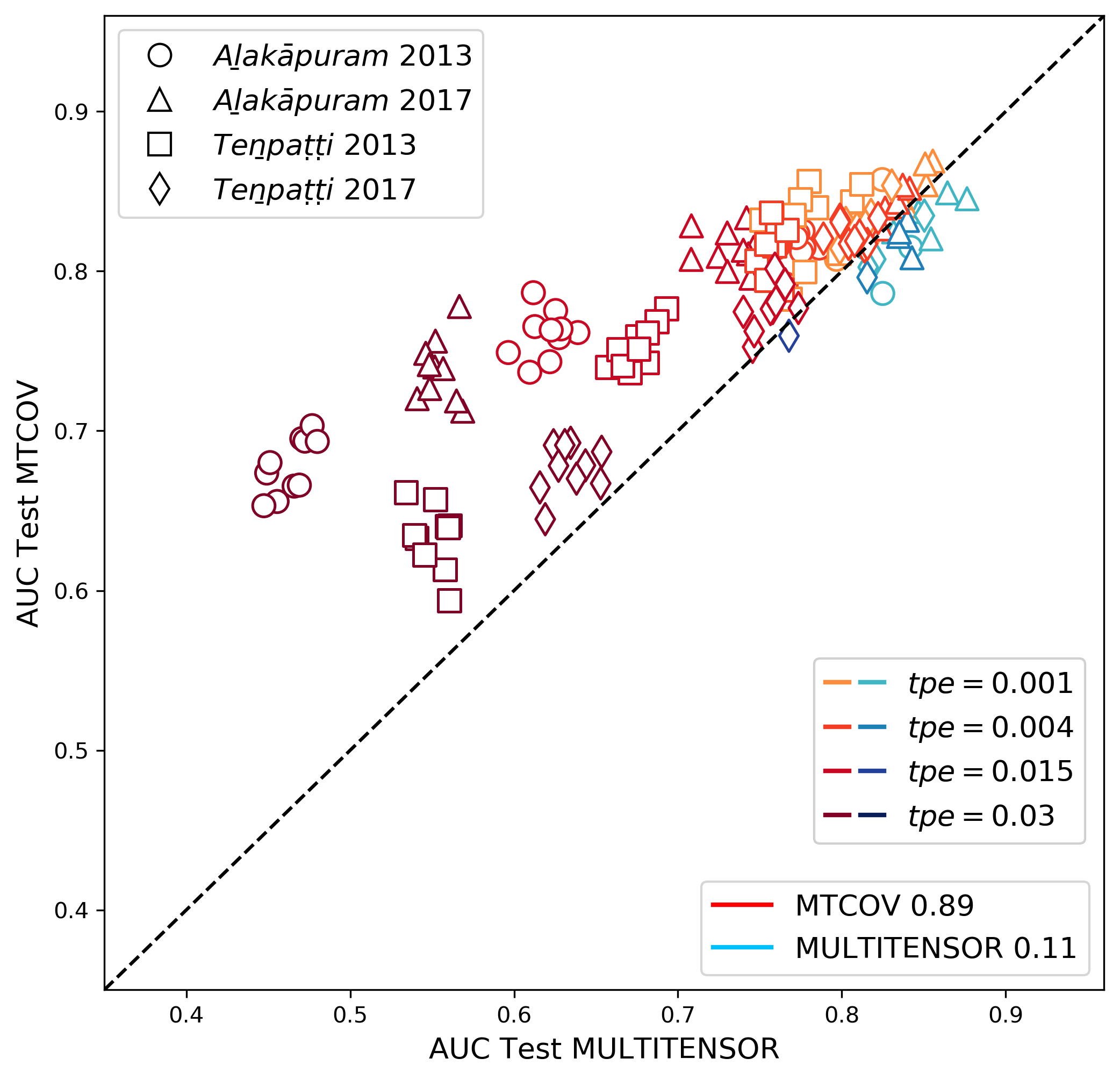}
  \end{minipage}\hfill
  \begin{minipage}[c]{0.5\linewidth}
    \caption{
 Trial-by-trial probabilistic link prediction with biased edge sampling. The values of AUC for \mtcov\text{}  and \mt\text{} are shown on the vertical axis and the horizontal axis respectively. The brightness represents the hardness of the settings in terms of biasing the edge sampling in the training. From bottom to top: $tpe=0.03$ (hard, dark color), $tpe=0.015$, $tpe=0.004$ (random),  $tpe=0.001$ (easy, light color). Points above the diagonal, shown in shades of red, are trials where \mtcov\text{}  is better performing than \mt\text{}. The fractions for which each method is superior are shown in the plot legend. We use the attribute Religion. }
 \label{fig: tpe1}
  \end{minipage}
\end{figure}

For validation, we split the dataset into training/test sets uniformly at random as explained in the Methods section. Table \ref{tab: randomsampling} reports the average results over ten runs for each network, and shows that \mtcov\text{} is capable of leveraging two sources of information to improve both performance metrics.
In fact, our algorithm systematically achieves the highest accuracy for attribute prediction and the highest AUC for edge prediction (boldface). While a good performance in attribute prediction is expected by design as we add this data into the model, the fact that it also boosts performance  in terms of edge prediction is not granted \textit{a priori}. Instead, it is a quantitative way to show that an attribute plays an important role in the
system. It also demonstrates the potential of capturing correlations between two different sources of information, which can have relevant applications, in particular when missing information of one kind. Notice in particular the improvement in AUC when using Caste compared to no attribute given (\mt).  The other attributes are less informative; in particular Age has a performance similar to \mt\text{} in edge prediction, signalling that it does not contribute to inform edge formation. Indeed, it has the smallest inferred $\gamma$ (always $< 0.5$), which gives also worse accuracy performance than the baseline, signalling again that this attribute may not be correlated with the community structure. All these results show the flexibility of  \mtcov \text{}  in adapting based on the data given in input: if warranted, it is able to ignore those attributes that are not correlated with network division and instead find communities that are mainly based on
the network structure.  Next, we test how adding node attributes impacts robustness against unbalanced data, where the ratio of positive examples (existing edges) observed in the training is different than that in the test set.
We denote the total probability of selecting an edge in the test as \textit{tpe} and consider values $\textit{tpe} \in \ccup{0.001,\, 0.004,\, 0.015, \, 0.03}$ denoting under-representation ($0.001$),  equal ($0.004$), and over-representation (values $0.015$ and $0.03$) compared to the uniform at random selection (empirically we find $tpe=0.004$). In these tests, we hold out $20\%$ of the entries of $A$ biasing their selection using the \textit{tpe} values; in addition, we give as input the whole design matrix $X$ (attributes) and measure link prediction performance. 
We observe that \mtcov\text{} is significantly more robust than the algorithm that does not use any attribute information, regardless of the value of $\gamma$. In fact, even though performance deteriorates as we decrease the number of positive examples in the training set (i.e., higher \textit{tpe}), \mtcov\text{} is less impacted by this, as shown in Fig.  \ref{fig: tpe2} (results reported in Supplementary Table S3). Notice in particular performance discrepancies when using the attribute Caste in the difficult regimes ($\textit{tpe} \in \ccup{0.015, 0.03}$): \mtcov's performance deteriorates only a little, while using the other attributes or no attribute makes performance significantly worse, with AUC down to 0.6 from a value higher than 0.8 for easier regimes. Moreover, notice that attributes with the same scaling parameter value can give different prediction results, underlying the necessity to consider both the value of the estimated $\gamma$ and the quality of the attribute to quantify its importance. This could explain why Caste provides always better results, given by the fact that its categories are more heterogeneous (i.e., more information) than Religion and Gender. The robustness of \mtcov\text{} is also confirmed by analyzing the performances on a trial-by-trail basis, each trial being a random sample of the held-out entries. As we show in Fig. \ref{fig: tpe1}, \mtcov\text{} better predicts links in $89\%$ of the trials and  never goes below the threshold of 0.5, the baseline random choice. These results demonstrate how adding another source of information helps when observing a limited amount of network edges. 
\begin{table}[h!]
\centering
\resizebox{0.9\textwidth}{!}{
\renewcommand\arraystretch{1.25}
\begin{tabular}{ll|cccc|cccc}
\clineB{1-10}{3}
\multicolumn{1}{c}{\textbf{}}  &\multicolumn{1}{c}{\textbf{}}        & \multicolumn{4}{c}{\textbf{ACCURACY for attribute prediction}}                                        & \multicolumn{4}{c}{\textbf{AUC for link prediction}}    \\
\multicolumn{1}{l}{\textbf{Attribute}} &\multicolumn{1}{l|}{\textbf{Method}} & \textbf{Ala 2013}           & \textbf{Ala 2017}           & \textbf{Ten 2013} & \textbf{Ten 2017}      & \textbf{Ala 2013}           & \textbf{Ala 2017}           & \textbf{Ten 2013}  & \textbf{Ten 2017}  \\ \hline
&\multicolumn{1}{l|}{\mt}  &&&&  & 0.771$\pm$0.009 & 0.835$\pm$0.006 &0.758$\pm$0.005&0.81$\pm$0.01\\ \hline
\multirow{3}{*}{Caste}       &
\multicolumn{1}{l|}{RP}   & 0.07 & 0.08 & 0.10& 0.11 & & & &\\ 
&\multicolumn{1}{l|}{MRF} & 0.556$\pm$0.009&  0.57$\pm$0.01  & 0.32$\pm$0.01 & 0.33$\pm$0.02& & & &\\ 
&\multicolumn{1}{l|}{\mtcov }  & \textbf{0.80$\pm$0.05} & \textbf{0.77$\pm$0.05} & \textbf{0.69$\pm$0.09} & \textbf{0.74$\pm$0.07}
& \textbf{0.837$\pm$0.009} & \textbf{0.858$\pm$0.008}&\textbf{0.829$\pm$0.006}&\textbf{0.82$\pm$0.01}\\ 
\hline
\multirow{3}{*}{Religion}   &\multicolumn{1}{l|}{RP}   & 0.33 & 0.33 & 0.50 & 0.50 & & & &\\ 
&\multicolumn{1}{l|}{MRF} & 0.837$\pm$0.008& 0.843$\pm$0.006 & 0.696$\pm$0.008 & 0.679$\pm$0.008 & & & &\\ 
&\multicolumn{1}{l|}{\mtcov }  & \textbf{0.96$\pm$0.02} & \textbf{0.95$\pm$0.03 } & \textbf{0.76$\pm$0.08} & \textbf{0.80$\pm$0.05}
&0.813$\pm$0.007 &0.83$\pm$0.01 &0.81$\pm$0.02& 0.80$\pm$0.01\\
\hline
\multirow{3}{*}{Age}   &\multicolumn{1}{l|}{RP}   & 0.09 & 0.08 & 0.09 & 0.08 & & & &\\ 
&\multicolumn{1}{l|}{MRF} & \textbf{0.135$\pm$0.005} &  \textbf{0.126$\pm$0.007}  & 0.126$\pm$0.005 & \textbf{0.128$\pm$0.008}& & & &\\ 
&\multicolumn{1}{l|}{\mtcov }  & 0.11$\pm$0.03 & 0.11$\pm$0.02 & \textbf{0.13$\pm$0.04} & 0.10$\pm$0.03
&0.80$\pm$0.01 & 0.823$\pm$0.008 &0.783$\pm$0.009 & 0.80$\pm$0.01\\
\hline
\multirow{3}{*}{Gender}   &\multicolumn{1}{l|}{RP}   & 0.50 & 0.50 & 0.50 & 0.50 & & & &\\ 
&\multicolumn{1}{l|}{MRF} & 0.584$\pm$0.009 &  0.58$\pm$0.01  & 0.56$\pm$0.01 & 0.55$\pm$0.01& & & &\\ 
&\multicolumn{1}{l|}{\mtcov }  & \textbf{0.61$\pm$0.05} & \textbf{0.65$\pm$0.04} & \textbf{0.58$\pm$0.08} & \textbf{0.71$\pm$0.08}
&0.79$\pm$0.02 & 0.831$\pm$0.009 &0.80$\pm$0.01 & 0.81$\pm$0.01\\
 \clineB{1-10}{3}
\end{tabular}}
\caption{Prediction performance on real multilayer networks with attributes. Results are averages and standard deviations over 10 independent trials of cross-validation with 80-20 splits selected uniformly at random (i.e., $\textit{tpe}=0.004$); the best performances are in boldface. Datasets are described in Table \ref{tab: Statistics}. RP is the performance of uniform random probability and MRF the one of the maximum relative frequency, see Methods section for details. 
}
\label{tab: randomsampling}
\end{table}
\begin{figure}[h!]
  \centering
    \includegraphics[width=0.8\linewidth]{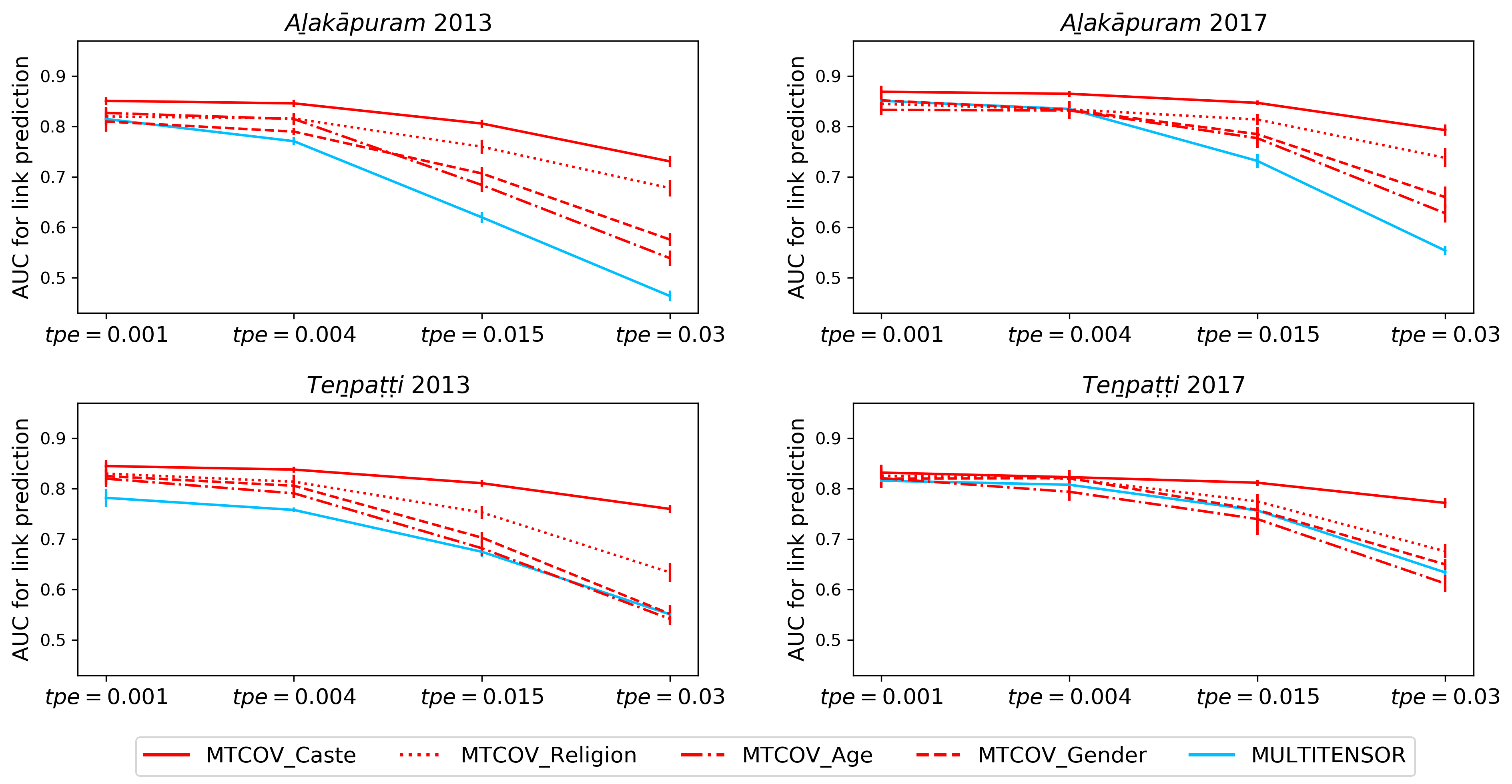}
\caption{Probabilistic link prediction with biased edge sampling. Results are AUC values of \mtcov\text{} and \mt\text{} on four social support networks in different held-out settings. Here $tpe$ indicates the total probability of selecting one edge (positive example) in the test set. We consider Caste, Religion, Age and Gender attributes;  results are averages and standard deviations over 10 independent runs. 
}
\label{fig: tpe2}
\end{figure}


\subsubsection*{Qualitative analysis of a social support network}
\begin{figure}[h!]
  \centering
    \includegraphics[width=1\linewidth]{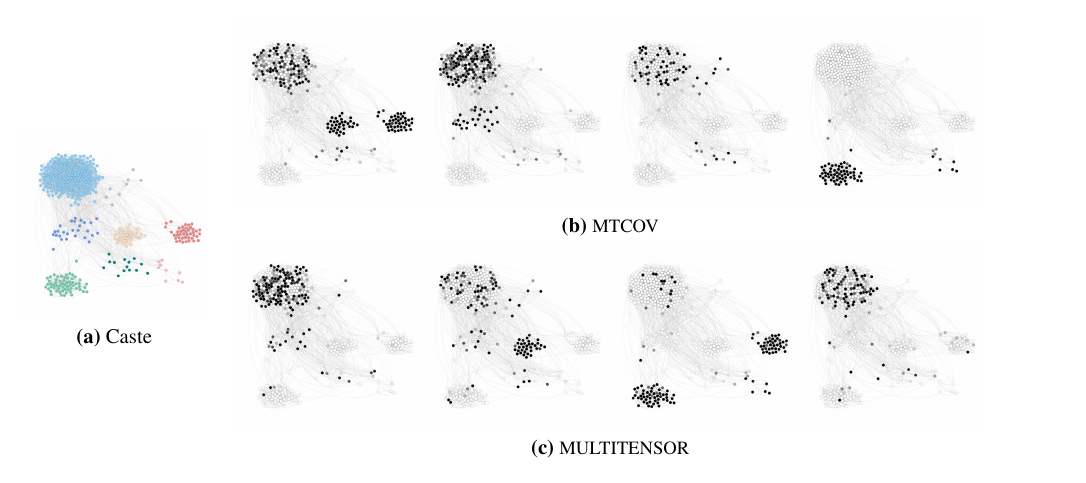}
\caption{Attributes and inferred communities. Nodes of the social support network of A\underline{l}ak\={a}puram in 2017 are colored by: (a) the attribute Caste (with colors as shown in Fig. \ref{fig: interpretability_Ala2017}); inferred communities by (b) \mtcov\text{} and (c) \mt\text{}. Darker values in the grey scales indicate higher values of the entry of the membership vector $u_i$. }
\label{fig:intNet}
\end{figure}

\begin{figure}[phbt]
  \begin{minipage}[c]{0.63\linewidth}
    \includegraphics[width=\linewidth]{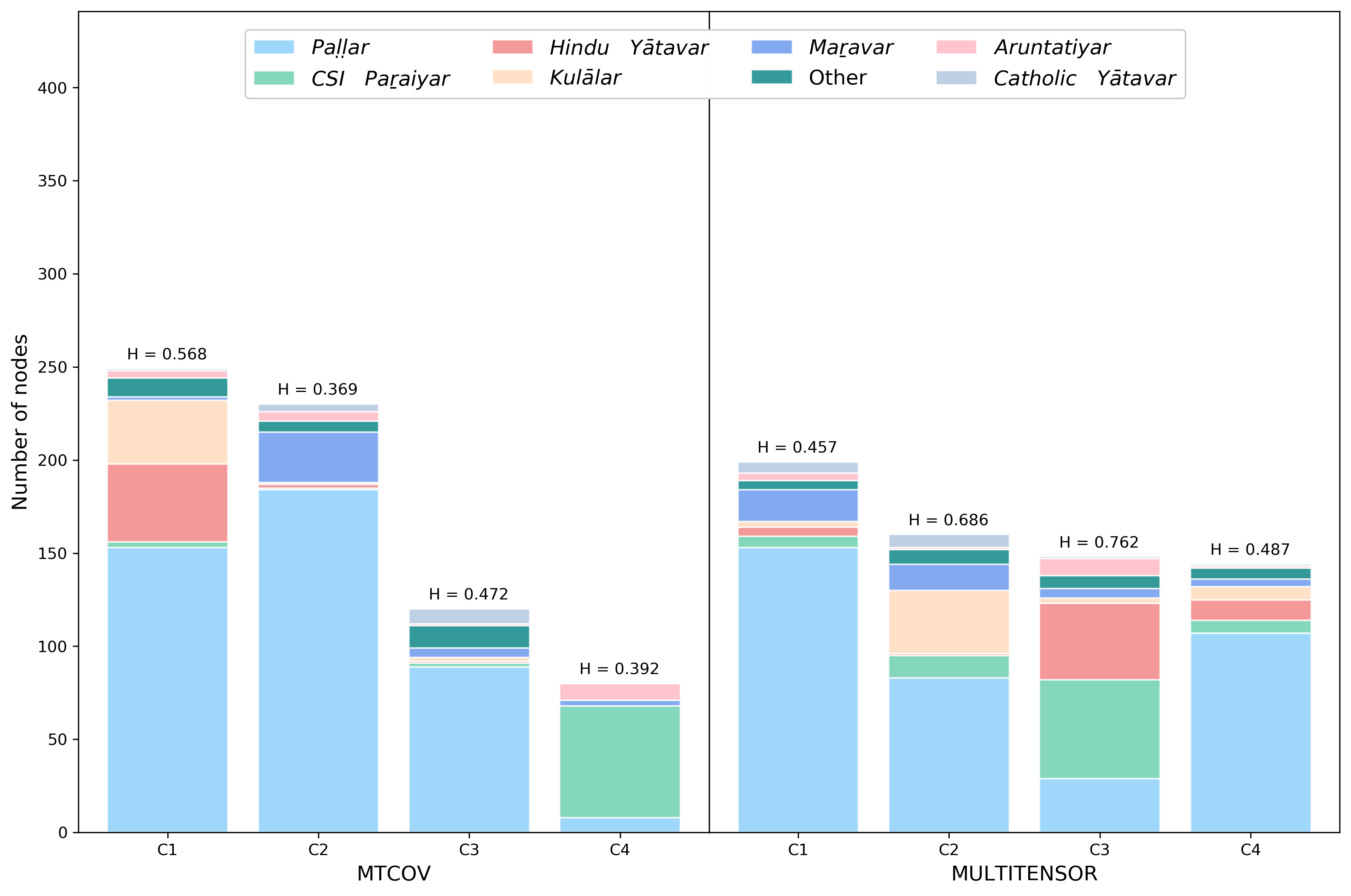}
  \end{minipage}\hfill
  \begin{minipage}[c]{0.35\linewidth}
    \caption{Partition of the attribute Caste inside each community detected by \mtcov\text{} and \mt\text{} in the social support network of A\underline{l}ak\={a}puram  in 2017. The category Other contains small categories having less than five individuals. The label on top of each bar is the value of the entropy of the variable Caste inside the corresponding community. Note that nodes can have mixed membership, here we build a group $k$ by adding to it all nodes $i$ that have a non-zero $k$-th entry $u_{ik}$. The number of nodes is $N=441$, corresponding to the maximum value of the y-axis plotted.}
\label{fig: interpretability_Ala2017}
\end{minipage}
\end{figure}

To demonstrate our  \mtcov\text{} model beyond prediction tasks and highlight its potential for interpretability, we show as an example its qualitative behavior on the real network of A\underline{l}ak\={a}puram  in 2017 (see Table \ref{tab: Statistics}). Specifically, we compare the communities extracted by our algorithm and those inferred by \mt. To ease comparison, we fix the same number of groups to $C=4$ for both algorithms and measure how caste membership distributes across communities, and fix $\gamma=0.8$ as obtained with cross-validation. Figure \ref{fig:intNet} shows the magnitude of each individual's inferred outgoing memberships $u_{i}$ in each group. While the communities identified by \mtcov \text{} and \mt \text{} show substantial similarities, \mtcov\text{} generally classifies castes more consistently into distinct communities, as we show in Fig. \ref{fig:intNet} and \ref{fig: interpretability_Ala2017}. To make a quantitative estimate of the different behaviors, we measure the entropy of the attribute inside each community $H_{k}=-\sum_{z=1}^{Z}f_{z}\log f_{z}/\log(Z)$, where $f_{z}$ is the relative frequency of the $z$-th caste inside a group $k$, and the denominator is the entropy of a uniform distribution over the $Z$ castes, our baseline for comparison. Values of $H_{k}$ close to 1 denote a more uniform distribution of castes, whereas smaller values denote an unbalanced distribution with most of the people belonging to a few castes. We find that \mtcov\text{} has smaller entropies over the groups, with two groups having the smallest values, 
 whereas \mt\text{} has the highest,
  showing its tendency to cluster individuals of different castes into the same group.
In addition, we observe that \mtcov\text{} has a more heterogenous group size distribution which seems to be correlated with Caste. Notably, the algorithms differ in how they	 place two caste groups that live in hamlets separated from the main village (the Hindu Y\= atavars and CSI Pa\underline{r}aiyars). With  \mt \text{}, they are grouped together, while with \mtcov \text{}, the Hindu Y\= atavars are joined up into a community with Pa\d l\d lars and Kul\= alars. While  \mt \text{} is clearly picking up the structural similarities of the two hamlets, this division makes little sense socially and culturally. In contrast, the way in which \mtcov \text{} defines a community which spans caste boundaries (\mtcov \text{} C1) aligns with ethnographic knowledge of the relations between these castes. Finally, we remark that there might be multiple meaningful community divisions in the network, and the fact that \mtcov's partition seems to better capture the distributions in the attribute Caste does not mean than one algorithm is better than the other. In fact, there might be other hidden topological properties that \mt's partition is picking up by being agnostic to caste membership. The choice of which algorithm to use should be made based on the final goal of the application at hand.


%



\subsection*{Results on single-layer networks} 
\begin{figure}[]
  \centering
    \includegraphics[width=0.7\linewidth]{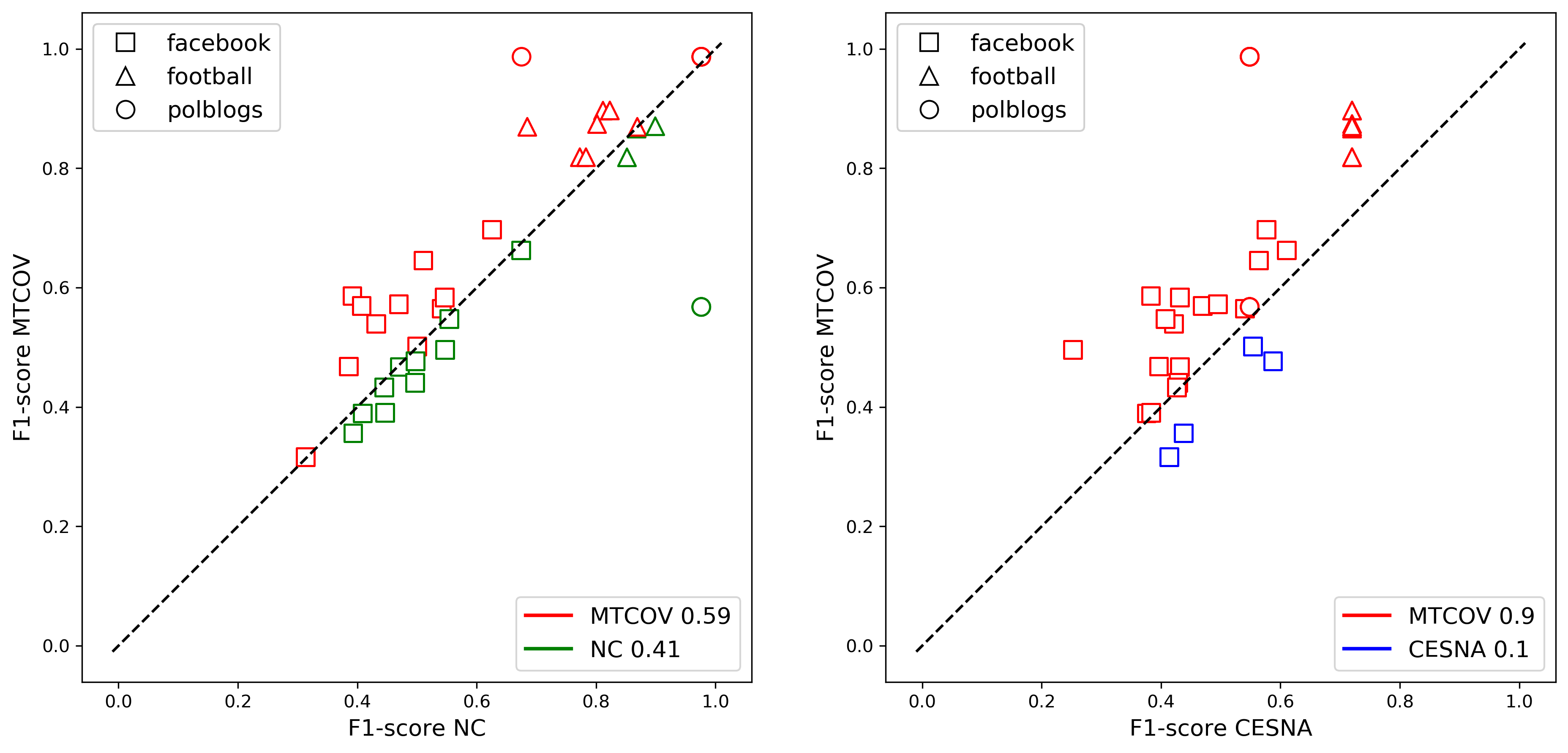}
\caption{Trial-by-trial performance in F1-score. We compare \mtcov\text{} on the y-axis, with on the x-axis (left) \nc\text{} and (right) \cesna. Markers denote the datasets: squares for \textit{facebook}, triangles for \textit{football} and circles for \textit{polblogs}. Points above the diagonal, shown in red, are trials where \mtcov\text{}  is more accurate than the other. The fractions for which each method is superior are shown in the plot legend. }
\label{fig: single-layer1}
\end{figure}

Our model can be used for single-layer networks as well. For these we can compare against two state-of-the-art algorithms, both probabilistic generative models but different in their assumptions: \cesna\text{} \cite{yang2013community} which considers overlapping communities and posits two independent Bernoulli distributions for network edges and node attributes; and  the model proposed by Newman and Clauset \cite{newman2016structure} (\nc) for non-overlapping communities, a Bayesian approach where the priors on the community memberships depend on the node attributes. \cesna,  similarly to our model, assumes conditional independence of the two likelihoods and introduces a regularization parameter between them; it uses block-coordinate ascent for parameters' estimation, while \nc\text{} uses an EM algorithm for parameters' estimation, similarly to what we do here. We test \mtcov\text{} against them on both synthetic and real single-layer networks with node attributes, with and without ground-truth. We transform directed networks to undirected because both \cesna\text{} and \nc\text{} do not distinguish for edge directionality. Results on synthetic data show that \mtcov\text{} and \nc\text{} have similar performance in correctly classifying nodes in their ground-truth communities and both are better than \cesna; the main difference is that \mtcov\text{} is more stable and has less variance for high attribute correlation,  in particular in the hard regime where classification is more difficult. We leave details in the Supplementary Section S4.  
For single-layer real networks, we use datasets with ground-truth candidates and node attributes: the ego-Facebook network (\textit{facebook}) \cite{facebook}, a set of 21 networks built from connections between a person's friends where potential ground-truth are circles of friends hand-labeled by the ego herself; the American College football network (\textit{football}) \cite{football}, a network of football teams playing against each other, where a ground-truth candidate is the conference to which each team belongs; and a network of political blogs (\textit{polblogs}) \cite{polblogs} where potential ground-truth communities are divided by \textit{left/liberal} and \textit{right/conservative} political parties, see Supplementary Section S4 for details. For each network, we run a 5-fold cross-validation procedure combined with grid-search for fixing the hyperparameter $\gamma$ (see ``Cross-validation tests and hyperparameter settings'' subsection for details; note that in this case we use the ground-truth value of $C$, hence $\gamma$ is the only hyperparameter left to be tuned). For \textit{facebook} we find that the average over the 21 networks is $\gamma=0.15$, which signals a low correlation between the covariates and the communities, whereas for the \textit{football} and \textit{polblogs} networks we obtain much higher values of $\gamma$ equal to $0.6$ and  $0.75$ respectively. \mtcov\text{} has better performance in terms of F1-score and Jaccard similarity across the majority of datasets, as shown in Table \ref{tab: single-layer}. This is also supported by a trial-by-trial comparison shown in Fig. \ref{fig: single-layer1} for F1-score (similar results are obtained for Jaccard), where we find that \mtcov\text{}  is more accurate in $59\%$ and $90\%$ of the cases than \nc\text{} and \cesna, respectively.

\begin{table}[]
\centering
\renewcommand{\arraystretch}{1.2}
{\begin{tabular}{lccc|ccc}
\clineB{1-7}{3}
\multicolumn{1}{c}{\textbf{}}        & \multicolumn{3}{c|}{\textbf{F1-score}}                                        & \multicolumn{3}{c}{\textbf{Jaccard similarity}}    \\
\multicolumn{1}{l|}{\textbf{Method}} & \textbf{facebook}           & \textbf{football}           & \textbf{polblogs} & \textbf{facebook}           & \textbf{football}           & \textbf{polblogs}  \\ \hline
\multicolumn{1}{l|}{\mtcov }           & \textbf{0.5$\pm$0.1} & \textbf{0.86$\pm$0.03} & 0.8$\pm$0.2   
& \textbf{0.4$\pm$0.1} & \textbf{0.82$\pm$0.04} & 0.8$\pm$0.2   \\
\multicolumn{1}{l|}{\nc}              & 0.48$\pm$0.08          & 0.82$\pm$0.06  &  \textbf{0.95$\pm$0.09}        & 0.36$\pm$0.08      & 0.75$\pm$0.08   &  \textbf{0.9$\pm$0.1}    \\
\multicolumn{1}{l|}{\cesna}           & 0.46$\pm$0.09          & 0.7$\pm$0.0          & 0.6$\pm$0.0           & 0.33$\pm$0.08          & 0.6$\pm$0.0             & 0.4$\pm$0.0   \\ 
\clineB{1-7}{3}
\end{tabular}}
\caption{Performance of methods \mtcov, \nc\text{} and \cesna\text{} on three datasets, according to two different measures used in the equation (\ref{eq: validation}). The results are averages and standard deviations over ten independent runs and the best outcomes are bolded. }
\label{tab: single-layer}
\end{table}

\section*{Discussion}
\label{sec: Discussion}
We present \mtcov, a generative model that performs overlapping community detection in multilayer networks with node attributes. We show its robustness in adapting to different scenarios, and its flexibility in exploiting the attributes that are more informative while ignoring those that are less correlated with the network communities. Our method is capable of estimating quantitatively the contribution given by the attributes  and incorporating them to improve prediction performance both in terms of recovering missing attributes and in terms of link prediction. This allows domain experts to investigate particular attributes and select relevant community partitions based on what type of node information they are interested in investigating. There are valuable possible extensions of this work. One example is to incorporate modeling of more complex data types for the attributes, for instance combinations of discrete and continuous attributes, or other types of extra information, like time-varying network elements, whether the attributes, node, edges or combinations of these. From a technical point of view, when the topological and attribute datasets are very unbalanced in size, this might impact their relative likelihood weight and thus inference. One should then consider automating the process of rescaling them accordingly, as a pre-processing step to be incorporated into the model. Similarly, hyperparameter selection would benefit from an automatized routine when more than one performance metric is considered. The relations between attributes and communities could be transferred across networks to predict missing information when having access to similar but incomplete datasets. We show examples of these here, where we studied two snapshots of the same village networks across time. While we leave these questions for future work, we provide an open source version of the code. 


\section*{Methods}
\label{sec: Methods}
\thispagestyle{empty}

We adapt recent ideas from the generative model behind \mt\text{}  \cite{Multitensor}, a multilayer mixed-membership model based on a Poisson tensor factorization \cite{Kolda2009}, to incorporate node attributes in a principled manner. 
 It can take in input directed and undirected networks, allowing different topological structures in each layer, including arbitrarily mixtures of assortative, disassortative and core-periphery structures. 
 We move beyond \mt\text{}  by incorporating node covariates via introducing a proper likelihood term that accounts for this extra information.
 We use the formalism of maximum likelihood estimation: we combine the structural and the node information into a global likelihood function and provide a highly scalable Expectation-Maximization algorithm for the estimation of parameters.
 
\subsection*{Model description and notation}
Consider a multilayer network of $N$ nodes and $L$ layers. This is a set of graphs $G = \{G^{(\alpha)}\left(\mathcal{V}, \mathcal{E}^{(\alpha)}\right)\}_{1\leq \alpha \leq L} $ defined on a set $\mathcal{V}$ of $N$ vertices shared across $L\geq1$ layers, and $\mathcal{E}^{(\alpha)}$ is the set of edges in the layer $\alpha$. Each layer $\alpha \in \{1,\dots, L\}$ is a graph $G^{(\alpha)}(\mathcal{V}, \mathcal{E}^{(\alpha)})$ with adjacency matrix $A^{(\alpha)} = [a_{ij}^{(\alpha)}] \in \mathbb{R}^{N\times N}$, where $a_{ij}^{(\alpha)}$ is the number of edges of type $\alpha$ from $i$ to $j$;  here we consider only positive discrete entries; for binary entries, $E=\sum_{i,j,\alpha} a_{ij}^{(\alpha)}$ is the total the number of edges.  Alternatively, we can consider a 3-way tensor $A$ with dimensions $N\times N \times L$.
 In addition, for each node $i\in \mathcal{V}$ consider the vector of covariates $X_i \in \mathbb{R}^{1\times K}$ (alternatively called also attributes or metadata), where $K$ is the total number of attributes. Here, for simplicity we focus on the case of $K=1$ and categorical covariates with $Z$ different categories. However, we can easily generalize to more than one covariate by encoding each possible combination of them as a different
value of one single covariate. For example, for two covariates being gender and nationality, we can encode $X_{i}$ being one covariate with possible values female/American, male/Spanish and so forth. 
One could also consider real-valued covariates by cutting them into bins. Nevertheless, a future expansion should include the possibility to work with any type of metadata.\\
 A community is a subset of vertices that share some properties. Formally, each node belongs to a community to an extent measured by a $C$-dimensional vector denoted \textit{membership}. Since we are interested in directed networks, for each node $i$ we assign two such vectors, $u_i$ and $v_i$ (for undirected networks we set $u =v$);  these determine how $i$ forms outgoing and incoming links respectively.
  Each layer $\alpha$ has an \textit{affinity} matrix $W^{(\alpha)} = [w_{kl}^{(\alpha)}]\in \mathbb{R}^{C \times C}$ which describes the density of edges between each pair ($k, l$) of groups. 
Each community $k\in \{1,\dots, C\}$ is linked to a category $z\in \{1,\dots, Z\}$ by a parameter $\beta_{kz}$, that explains how much information of the $z$-th category is used to create the $k$-th community. To summarize, we consider two types of observed data: the adjacency tensor $A = \{A^{(\alpha)}\}_{1\leq \alpha \leq L}$ and the design matrix $X=\{X_i\}_{i\in \{1,\dots, N\}}$; the first contains information about the networks topology structure, the latter about the node covariates. In addition, we have the model parameters that we compactly denote as $\Theta=\ccup{U,V,W,\beta}$.

\begin{figure}[htp]%
\centering
\includegraphics[width=0.6\linewidth]{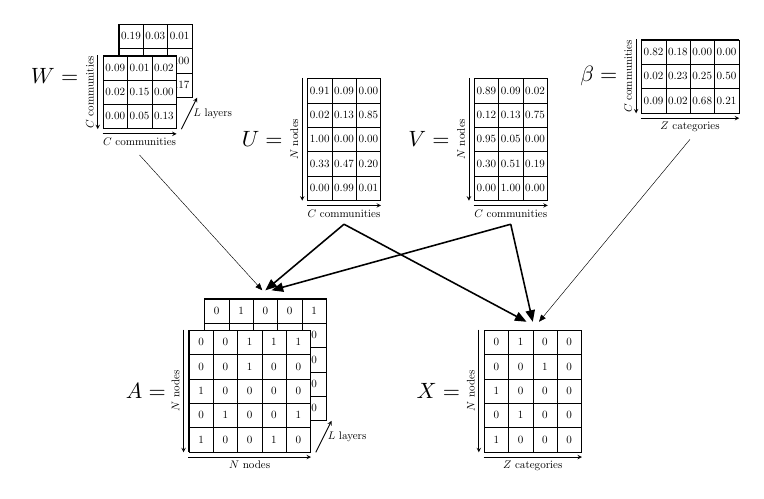}
 \caption{Graphical model representation of the algorithm \mtcov\text{}. $A$ is the adjacency tensor, $X$ is the design matrix and $W, U, V, \beta$ are the latent parameters $\Theta$. The membership matrices $U$ and $V$ couple the two datasets, and this is highlighted by the stronger thickness; whereas $W$ and $\beta$ are specific to the adjacency tensor and design matrix respectively. Here we present an example with binary adjacency matrix $A$, but the model is valid for more general weighted networks. }     
 \label{fig:graphicalmodel}%
\end{figure}

The goal is to find the latent parameters $\Theta$ using the data $A$ and $X$. In other words, 
given an observed multilayer network with adjacency tensor $A$ and design matrix $X$, our goal is to simultaneously infer the node's membership vectors $u_i$ and $v_i$ $\forall i\in \{1,...,N \}$; the affinity matrices $W^{(\alpha)}$, $ \forall \alpha \in \{1,\dots, L\}$, and the matrix $\beta = [\beta_{kz}] \in \mathbb{R}^{C \times Z}$, which captures correlations between communities and attributes. A visual overview of the proposed model is shown in Fig. \ref{fig:graphicalmodel}. 
We consider a probabilistic generative model where \mtcov\text{} generates the network and the attributes probabilistically, assuming an underlying structure consisting of $C$ overlapping communities. We adopt a maximum likelihood approach where, given the latent parameters $\Theta$, we assume that the data $A$ and $X$ have independent likelihoods;
in other words, we assume that $A$ and $X$ are \textit{conditionally independent} given the latent parameters $\Theta$. In addition, we assume that the memberships $U$ and $V$ couple the two datasets, as they are parameters shared between the two likelihoods; whereas the $W$ and $\beta$ are specific to the adjacency and design matrix respectively.
 We describe separately the procedures for modeling the topology of the network and the node attributes and then we show how to combine them in a unified log-likelihood framework. 

\subsection*{Modeling the network topology}
In modeling the likelihood of the network topology, we adopt the ideas behind \mt: we assume that the expected number of edges of type $\alpha$ from $i$ to $j$ is given by the parameter:
\begin{align}\label{eqn:meanPoisson}
M_{ij}^{(\alpha)}= \sum_{k,l =1}^C u_{ik}v_{jl}w_{kl}^{(\alpha)} .
\end{align}
We then assume that each entry $a_{ij}^{(\alpha)}$ of the adjacency tensor is extracted from a Poisson distribution with parameter $M_{ij}^{(\alpha)}$. This is a common choice for network data \cite{ball2011,gopalan2013,gopalan-rec} as it leads to tractable and efficient algorithmic implementations, compared for instance with other approaches that use Bernoulli random variables \cite{yang2013community,newman2016structure}; it also allows the flexibility of treating both binary and integer-weighted networks. We further assume that, given the memberships and affinity matrices, the edges are distributed independently;  this is again a conditional independence assumption. 

We can then write the likelihood of the network topology as:
\begin{align}\label{eqn:poisson}
P_{G}(A|U,V, W) = \prod_{i,j =1}^N  \prod_{\alpha =1}^L   \frac{{e}^{-M_{ij}^{(\alpha)}} \left(M_{ij}^{(\alpha)}\right)^{A_{ij}^{(\alpha)}}}{A_{ij}^{(\alpha)} !} ,
\end{align}

which leads to the log-likelihood $\mathcal{L}_G(U,V,W)$ for the \textit{structural dimension}:
\be\label{eqn:l_structural}
\mathcal{L}_G(U,V,W) = \sum_{i,j,\alpha}  \bigg[ A_{ij}^{(\alpha)}  \text{log} \sum_{k,l} u_{ik}v_{jl}w_{kl}^{(\alpha)} - \sum_{k,l} u_{ik}v_{jl}w_{kl}^{(\alpha)} \bigg],
\ee
where we have neglected constants that do not depend on the parameters.
\subsection*{Modeling the node attributes}

In modeling the likelihood of the attributes, we assume that this extra information is generated from the membership vectors; this captures the intuition that knowing a node's community membership helps in predicting the value of the node's attribute. This assumption has also been made in other models for single-layer attributed networks \cite{yang2013community} where one wants to enforce the tendency that nodes in the same community (for assortative structures) are likely to share common
attributes. Different approaches \cite{ airoldi2011confidence,sweet2015incorporating} assume instead independence between attributes and membership, which follows a different idea of observing an interaction between individuals if either they belong to the same community (for assortative structures) or they share an attribute or both. \\
 Then, we model the probability of observing the $z$-th category for the attribute covariate of node $i$ as the parameter:
\begin{align}
\label{eq: piz}
\pi_{iz} = \frac{1}{2}\sum_{k=1}^C \beta_{kz} (u_{ik}+v_{ik}) ,
\end{align}
where $\beta_{kz}$ is the probability of observing a particular category $z$ together with a community $k$; thus $\pi_i= (\pi_{i1} \dots, \pi_{iZ})$ is a $Z$-dimensional vector such that $ \pi_{iz} \in \rup{0,1}$ and $\sum_{z=1}^Z \pi_{iz} =1,  \forall i$. 
For convenience, we consider one-hot encoding for $x_i = (x_{i1}, \dots, x_{iZ})$, the realization of the random variable $X_i$: $x_{iz}=1$ if node $i$ has attribute corresponding to category $z$, $0$ otherwise and $\sum_{z=1}^Z x_{iz} =1$; the original design matrix $X_{N\times1}$ is thus translated into a binary matrix $X_{N\times Z}$.\\
We then assume that each entry $X_{i}$ of the design matrix is extracted from a multinomial distribution of parameter $\pi_{i}$, which yields the likelihood of the covariates:
\begin{align} \label{eqn:multinomial}
P_{X}(X_i=x_i|U,V,\beta) = P_{X}(X_{i1}=x_{i1}, \dots, X_{iZ}= x_{iZ}|U,V,\beta) =  \pi_{i1}^{x_{i1}} \dots  \pi_{iZ}^{x_{iZ}}\ \ .
\end{align}

In order to satisfy the sum constraint $\sum_{z=1}^Z \pi_{iz} =1$, we impose the normalizations $\sum_{z=1}^Z \beta_{kz} =1$, valid $\forall k$ and $\sum_{k=1}^C  u_{ik}= \sum_{k=1}^C v_{ik} =1$, valid $\forall i$. 
Such constraints are a particular case for which the general constraint for the multinomial parameter is satisfied. Although they are not the only choices, they allow us to give a probabilistic meaning to the components of $\beta$ and the memberships $U$ and $V$.
As done for the network's edges, we assume conditional independence for the attributes on the various nodes.
This leads to the log-likelihood $\mathcal{L}_{X}(U,V,\beta)$ for the \textit{attribute dimension}: 
\be \label{eq:l_covariate}
\mathcal{L}_{X}(U,V,\beta) = \sum_{i=1}^N \sum_{z=1}^Z x_{iz} \, \text{log}(\pi_{iz})
=\sum_{i, z} x_{iz} \, \text{log}\bigg(\frac{1}{2}\sum_{k}\beta_{kz} (u_{ik}+v_{ik})\bigg) .
\ee

Note, we assume that the attributes have values that can be binned in a finite number $Z$ of unordered categories and the attributes do not need to be one-dimensional. Indeed, we can encode each combination of more attributes as a different value of one-dimensional ``super-attribute''. The model will not be affected, but the computational complexity might increase. 

\subsection*{Inference with the EM algorithm}
Having described how the model works and its main assumptions and intuitions, we now turn our attention to describe how to fit the parameters to the data, in other words, how to perform inference. 
We assume conditional independence between the network and attribute variables, thus we can decompose the total log-likelihood into a sum of two terms $\mathcal{L}(U,V,W,\beta)=\mathcal{L}_G(U,V,W)+ \mathcal{L}_{X}(U,V,\beta)$. However, in practice, we can improve parameters' inference performance by better balancing the contributions of the two terms as their magnitude can be on different scales, thus the risk of biasing the total likelihood maximization towards one of the two terms. For this, we introduce a scaling parameter $\gamma \in [0,1]$ that explicitly controls the relative contribution of the two terms.
The total log-likelihood is then:
\bea\label{eqn:totalL}
\mathcal{L}(U,V,W,\beta)&=&\bup{1- \gamma}\mathcal{L}_G(U,V,W)+ \gamma \, \mathcal{L}_{X}(U,V,\beta)\\
&=&(1-\gamma) \sum_{i,j,\alpha}  \bigg[A_{ij}^{(\alpha)}  \text{log} \sum_{k,l} u_{ik}v_{jl}w_{kl}^{(\alpha)} - \sum_{k,l} u_{ik}v_{jl}w_{kl}^{(\alpha)}\bigg] 
+\gamma \,  \sum_{i, z} x_{iz} \text{log}\bigg(\frac{1}{2}\sum_{k} \beta_{kz} (u_{ik}+v_{ik})\bigg) \nonumber
\eea


 Varying $\gamma$ from 0 to 1 lets us interpolate between two extremes: analyzing the data purely in terms of the network topology or purely in terms of the attribute information. One can either fix this \textit{a priori} based on the goal of the application, closer to 0 for link prediction or closer to 1 for attribute classification, or this can be treated as a hyperparameter that must be estimated, whose optimal value is obtained by fitting the data \textit{via} tuning techniques (for instance cross-validation). This approach provides a natural quantitative measure for the dependence between the communities and the two sources of information. Notice that one can rescale \textit{a priori} each likelihood term individually in order to control even more their magnitudes, and then add it to equation (\ref{eqn:totalL}). This choice should be made based on the dataset at hand. Here we consider rescaling $\mathcal{L}_G$ and $\mathcal{L}_X$ only in studying the social support networks of Indian villages, as we have enough data for estimating the normalization coefficients; see Supplementary Section S3 (A) for details.

We wish to find the $\Theta = ({U,V, W,\beta})$ that maximizes equation (\ref{eqn:totalL}).  In general, this is computationally difficult, but we make it tractable by adopting a variational approach using an Expectation-Maximization (EM) algorithm \cite{dempster1977maximum}, similar to what done by De Bacco et al. \cite{Multitensor}, but extended here to include attribute information. Namely, we introduce two probability distributions: $h_{ikz}$ and $\rho_{ijkl}^{(\alpha)}$. For each $i, z$ with $X_{iz}=1$, $h_{izk}$ represents our estimate of the probability that the $i$-th node has the $z$-th category, given that it belongs to the community $k$. On the other hand, for each $i, j, \alpha$ with $A_{ij}^{(\alpha)} =1$, $\rho_{ijkl}^{(\alpha)}$ is the probability distribution over pairs of groups $k, l$. 

Using Jensen's inequality $\log \bar{x} \geq \overline{\log x}$ for each log-likelihood term gives: 
\begin{align}
\mathcal{L}_X(U,V,\beta)  &\geq \sum_{i, z} x_{iz} \sum_{k} h_{izk}\log \frac{\beta_{kz} (u_{ik}+v_{ik})}{2h_{izk}}  
=\sum_{i,z,k} x_{iz} \rup{h_{izk}\,\log \beta_{kz} (u_{ik}+v_{ik}) - h_{izk}\, \log 2h_{izk}} = \mathcal{L}_X(U,V,\beta, h) \label{eq:lxh} \\
\mathcal{L}_G(U,V,W) &\geq \sum_{i,j,k,l,\alpha}\bigg[ A_{ij}^{(\alpha)}\bigg(\rho_{ijkl}^{(\alpha)} \, \log u_{ik}v_{jl}w_{kl}^{(\alpha)} - \rho_{ijkl}^{(\alpha)}\log \rho_{ijkl}^{(\alpha)}\bigg) - u_{ik}v_{jl}w_{kl}^{(\alpha)}\bigg] = \mathcal{L}_G(U,V,W, \rho) \label{eq:lgp} \quad.
\end{align}


These lower bounds hold with equality when
\begin{align} \label{eqn:Estep}
h_{izk}=\frac{\beta_{kz} (u_{ik}+v_{ik})}{\sum_{k'} \beta_{k'z} (u_{ik'}+v_{ik'})} \quad,\quad \rho_{ijkl}^{(\alpha)} =\frac{u_{ik}v_{jl}w_{kl}^{(\alpha)}}{\sum_{k',l'} u_{ik'}v_{jl'}w_{k'l'}^{(\alpha)}} \quad,
\end{align}	
thus maximizing $\mathcal{L}_X(U,V,\beta)$ is equivalent to maximizing $\mathcal{L}_X(U,V,\beta, h)$; similarly for $\mathcal{L}_G(U,V,W)$ and $ \mathcal{L}_G(U,V,W, \rho)$ (this was also the same result derived by De Bacco et al. \cite{Multitensor}). Overall, we aim at maximizing $\mathcal{L}(U,V,W,\beta, h,\rho)=\bup{1- \gamma}\mathcal{L}_G(U,V,W, \rho)+ \gamma \, \mathcal{L}_{X}(U,V,\beta,h)$, in analogy with what was done before.  
These maximizations can be performed by alternatively updating a set of parameters while keeping the others fixed. The EM algorithm performs these steps by alternatively updating $h$, $\rho$ (Expectation step) and $\Theta$ (Maximization step); this is done starting from a random configuration until $\mathcal{L}(\Theta, h,\rho)$  reaches a fixed point. 
Calculating equation (\ref{eqn:Estep}) represents the E-step of the algorithm. The M-step is obtained by computing partial derivatives of $\mathcal{L}(\Theta, h,\rho)$ with respect to the various parameters in $\Theta$ and setting them equal to zero. We add Lagrange multipliers $\lambda=\bup{\lambda^{(\beta)},\lambda^{(u)}, \lambda^{(v)}}$ to enforce constraints:
\be  \label{eqn:LBeta}
\mathcal{L}^{'}(\Theta, h,\rho,\lambda)= \mathcal{L}(\Theta, h,\rho) - \sum_{k} \lambda^{(\beta)}_{k} \bup{\sum_{z=1}^Z \beta_{kz} -1} - \sum_{i} \lambda^{(u)}_{i} \bup{\sum_{k=1}^C u_{ik} -1} - \sum_{i} \lambda^{(v)}_{i} \bup{\sum_{k=1}^C v_{ik} -1} .
\ee


For instance, focusing on the update for $\beta_{zk}$,
setting the derivative with respect to it in equation (\ref{eqn:LBeta}) to zero and enforcing the constraint $\sum_{z=1}^Z \beta_{kz} =1$ gives $\lambda^{(\beta)}_k = \gamma \sum_{i,z}  x_{iz} h_{izk}$; plugging this back 
 finally gives:
\begin{align}
\label{eqn:beta}
    \beta_{kz} = \frac{\sum_{i} \,x_{iz} \,h_{izk}}{\sum_{i,z}  \,x_{iz} \,h_{izk}} \quad ,
\end{align}
which is valid for $\gamma \neq 0$.
Doing the same for the other parameters yields (see Supplementary Section S1 for details):
\bea
  u_{ik} &=& \frac{\gamma \, \sum_z \,x_{iz} h_{izk} + (1-\gamma)\, \sum_{j,l,\alpha}A_{ij}^{(\alpha)} \rho_{ijkl}^{(\alpha)}}{ \gamma + (1-\gamma) \sum_{j,\alpha}A_{ij}^{(\alpha)}}\label{eqn:u} \\
  v_{ik}&=&    \frac{\gamma \,\sum_z\, x_{iz} h_{izk} + (1-\gamma)\, \sum_{j,l,\alpha}A_{ji}^{(\alpha)} \rho_{jilk}^{(\alpha)}}{ \gamma + (1-\gamma) \sum_{j,\alpha}A_{ji}^{(\alpha)}} \label{eqn:v} \\
  w_{kl}^{(\alpha)} &=& \frac{\sum_{i,j} A_{ij}^{(\alpha)} \rho_{ijkl}^{(\alpha)}}{\sum_i u_{ik} \sum_j v_{jl}} \label{eqn:w}\quad , 
\eea
where equation (\ref{eqn:w}) is valid for $\gamma \neq 1$.
The EM algorithm thus consists in randomly initializing the parameters $\Theta$ and then repeatedly alternating between updating $h$ and $\rho$ using equation (\ref{eqn:Estep}) and updating $\Theta$ using equations (\ref{eqn:beta})-(\ref{eqn:w}) until $\mathcal{L}(\Theta, h,\rho)$ reaches a fixed point. A pseudo-code  is given in Algorithm \ref{alg:EM}. In general, the fixed point is a local maximum but we have no guarantees that it is also the global one. In practice, we run the algorithm several times, starting from different random initializations and taking the run with the largest final $\mathcal{L}(\Theta, h,\rho)$. 
The computational complexity per iteration scales as $O(M\,C^{2}+NCZ)$, where $M$ is the total number of edges summed across layers. In practice, $C$ and $Z$ have similar order of magnitude, usually much smaller than the system size $M$; for sparse networks, as is often the case for real datasets, $M \propto N$, thus the algorithm is highly scalable with a total running time linear in the system size.  An experimental analysis of the computational time is provided in the Supplementary Section S2.\\
Notice that, although we started from a network log-likelihood $\mathcal{L}_{G}(U,V,W)$ similar to the one proposed in the \mt\text{}  model \cite{Multitensor}, the only update preserved from that is the one of $w_{kl}$ in equation (\ref{eqn:w}). The updates for $  u_{ik}$ and $v_{ik}$ are instead quite different; the main reason is that here we incorporated the node attributes, which appear both explicitly and implicitly (through $h$) inside the updates. In addition, here we enforce normalizations like $\sum_{k}u_{ik}=1$, not enforced in \mt . This implies that our model restricted to $\gamma=0$, i.e., no attribute information, does not correspond exactly to \mt . This also implies that, upon convergence, we can directly interpret the memberships as \textit{soft} community assignments (or overlapping) without the need of post-processing their values; in words, $u_{ik}$ represent the probability of node $i$ to belong to the \textit{outgoing} community $k$, similarly for $v_{ik}$ and an \textit{incoming} membership. This distinction is necessary when considering directed networks. If one is interested in recovering \textit{hard} memberships, where a node is assigned to only one community, then one can choose the community corresponding to the maximum entry of $u$ or $v$.

\setlength{\textfloatsep}{5pt}
\begin{algorithm}[H]
	 \setstretch{0.7}
	\DontPrintSemicolon \; \KwIn{network $A=\{A_{ij}\}_{i,j=1}^{N}$, design matrix $X=\ccup{x_{iz}}_{i=1}^{N}$, number of communities $C$, hyperparameter $\gamma$}
	\BlankLine
	\KwOut{membership vectors $U=\rup{u_{ik}},\, V=\rup{v_{ik}}$; network-affinity matrix $W=\rup{w_{kl}}$; attribute-affinity matrix $\beta=\rup{\beta_{kz}}$.}
	
	\BlankLine
	 Initialize $U,V,W,\beta$ at random. 
	 \BlankLine
	 Repeat until convergence:
	 \BlankLine
	\quad 1. Calculate $h$ and $\rho$ (E-Step): 
	\be
	h_{izk}=\frac{\beta_{kz} (u_{ik}+v_{ik})}{\sum_{k'} \beta_{k'z} (u_{ik'}+v_{ik'})} 
	\quad,\quad \rho_{ijkl}^{(\alpha)} =\frac{u_{ik}v_{jl}w_{kl}^{(\alpha)}}{\sum_{k',l'} u_{ik'}v_{jl'}w_{k'l'}^{(\alpha)}} \nonumber
	\ee
	
	 \quad 2. Update parameters $\Theta$ (M-Step):  
	\BlankLine
	\quad \quad \quad 
		i) for each node $i$ and community $k$ update memberships:
		\bea
		\quad  u_{ik} &=& \frac{\gamma \, \sum_z \,x_{iz} h_{izk} + (1-\gamma)\, \sum_{j,l,\alpha}A_{ij}^{(\alpha)} \rho_{ijkl}^{(\alpha)}}{ \gamma + (1-\gamma) \sum_{j,\alpha}A_{ij}^{(\alpha)}} \nonumber\\
\quad  v_{ik}&=&    \frac{\gamma \,\sum_z\, x_{iz} h_{izk} + (1-\gamma)\, \sum_{j,l,\alpha}A_{ji}^{(\alpha)} \rho_{jilk}^{(\alpha)}}{ \gamma + (1-\gamma) \sum_{j,\alpha}A_{ji}^{(\alpha)}} \nonumber
		\eea
	\quad \quad \quad
		ii) if $\gamma \neq 1$, for each pair of communities $(k,l)$ update network-affinity matrix:
		\be
		\quad w_{kl}^{(\alpha)} = \frac{\sum_{i,j} A_{ij}^{(\alpha)} \rho_{ijkl}^{(\alpha)}}{\sum_i u_{ik} \sum_j v_{jl}} \nonumber
		 \ee
	\quad \quad \quad
		iii) if $\gamma \neq 0$, for each pair of community-attribute $(k,z)$ update attribute-affinity matrix:
		\be
		\quad \beta_{kz} = \frac{\sum_{i} \,x_{iz} \,h_{izk}}{\sum_{i,z}  \,x_{iz} \,h_{izk}}  \nonumber
		 \ee

	\caption{\mtcov - EM algorithm}
	\label{alg:EM}
\end{algorithm}

\subsection*{Evaluation metrics}
We adopt two different criteria for performance evaluation, based on having or not having access to ground-truth values for the community assignments. The first case applies to synthetic-generated data, the second to both synthetic and real-world data. We explain performance metrics in detail below.

\paragraph{Ground-truth available} In the presence of a known partition, we measure the agreement between the set of ground-truth communities $\mathcal{C}^*$ and the set of detected communities $\mathcal{C}$ using metrics for recovering both hard and soft assignments. For hard partitions, the idea is to match every detected community with its most similar ground-truth community and measure similarity $\delta(\mathcal{C}_i^*, \mathcal{C}_j)$ (and vice versa for every ground-truth community matched against a detected community) as done by Yang et al. \cite{yang2013community}. The final performance is the average of these two comparisons:
\begin{equation}
    \frac{1}{2|\mathcal{C}^*|} \sum_{\mathcal{C}_i^*\in \mathcal{C}^*} \max_{\mathcal{C}_j\in \mathcal{C}} \delta(\mathcal{C}_i^*, \mathcal{C}_j) + 
    \frac{1}{2|\mathcal{C}|} \sum_{\mathcal{C}_j\in \mathcal{C}} \max_{\mathcal{C}_i^*\in \mathcal{C}^*} \delta(\mathcal{C}_i^*, \mathcal{C}_j) \quad,
   \label{eq: validation}
\end{equation}

where here we consider as similarity metric $\delta(\cdot)$ the F1-score and the Jaccard similarity.
In both cases, the final score is a value between 0 and 1, where 1 indicates the perfect matching between detected and ground-truth communities. 
For soft partitions, we consider two standard metrics for measuring distance between vectors as done by De Bacco et al. \cite{Multitensor}, such as \textit{cosine similarity} (CS) and $L_{1}$ error, averaged over the nodes:
\bea
CS(U,U^{0})&=& \f{1}{N} \sum_{i=1}^{N} \f{u_{i} \cdot u^{0}_{i}}{||u_{i}||_{2}\,||u^{0}_{i}||_{2}} =\f{1}{N} \sum_{i=1}^{N} \sum_{k=1}^{C} \f{u_{ik} \, u^{0}_{ik}}{||u_{i}||_{2}\,||u^{0}_{i}||_{2}} \\
L_{1}(U,U^{0})&=& \f{1}{2N} \sum_{i=1}^{N}|| u_{i}-u_{i}^{0}||_{1}=\f{1}{2N} \sum_{i=1}^{N}\sum_{k=1}^{C}| u_{ik}-u_{ik}^{0}| \quad,
\eea
where $u_{i}$ is the $C$-dimensional vector containing the $i$-th row of $U$, representing the detected membership and similarly for $u_{i}^{0}$ for the ground-truth $U^{0}$. The factor $1/2$ ensures that the $L_{1}$ distance ranges from 0 for identical distributions to 1 for distributions with disjoint support. 
Similarly to the what done for hard partitions, we match the ground-truth and detected communities by choosing the permutation of $C$ groups that gives the highest cosine similarity or smallest $L_{1}$ distance. 

\paragraph{Ground-truth not available} In the absence of ground-truth, these metrics cannot be computed, and one must resort to other approaches for model evaluation. Here we consider performance in prediction tasks when hiding part of the input datasets while fitting the parameters, and in particular on the extent to which partial knowledge  of network edges helps predict node attributes and vice versa.  Thus we consider a measure for link-prediction and one for correct retrieval of the attributes. 
For link-prediction, we used the AUC statistic, equivalent to the area under the receiver-operating characteristic (ROC) curve \cite{hanley1982meaning}. It represents the probability that a randomly chosen missing connection (a true positive) is given a higher score than a randomly chosen pair of unconnected vertices (a true negative). Thus, an AUC statistic equal to $0.5$ indicates random chance, while the closer it is to $1$, the more our model's predictions are better than chance. We measure the probability of observing an edge as the predicted expected Poisson parameters of equation (\ref{eqn:meanPoisson}). For the attribute, instead, we use the accuracy as a quality measure. For each node, we compute the predicted expected multinomial parameter $\pi_i$ using equation (\ref{eq: piz}). We then assign to each node the category with the highest probability, computing the accuracy as the ratio between the correctly classified examples over the total number of nodes. As baselines, we compare with the accuracy obtained with a random uniform probability and the highest relative frequency observed in the training set. 

\subsection*{Cross-validation tests and hyperparameter settings}
We perform prediction tasks using cross-validation with 80-20 splits: we use $80\%$ of the data for training the parameters and then measure AUC and accuracy on the remaining $20\%$ test set. Specifically, for the network topology, we hold out $20\%$ of the triples $(i,j,\alpha)$; for the attributes, we hold out $20\%$ of the entries of the categorical vector.\\
Our model has two hyperparameters, the scaling parameter $\gamma$ and the number of communities $C$. We estimate them by using $5$-fold cross-validation along with grid search to range across their possible values. We then select the combination ($\hat{C},\hat{\gamma}$) that returns the best average performance over the cross-validation runs. Standard cross-validation considers performance in terms of a particular metric. However, here we have two possible ones which are qualitatively different, i.e., AUC and accuracy. Depending on the task at hand, one can define performance as a combination of the two, bearing in mind that the values of ($\hat{C},\hat{\gamma})$ at the maximum of either of them might not coincide.  Here we select ($\hat{C},\hat{\gamma})$ as the values are jointly closer to both the maximum values.  In the experiments where one of the two hyperparameters is fixed \textit{a priori}, we run the same procedure but vary with grid search only the unknown hyperparameter.


\section*{Data availability}
The code used for the analysis and to generate the synthetic data is publicly available and can be found at \url{https://github.com/mcontisc/MTCOV}.

\section*{Acknowledgements}
This work was partially supported by the Cyber Valley Research Fund. The authors thank the International Max Planck Research School for Intelligent Systems (IMPRS-IS) for supporting Martina Contisciani. The authors are grateful for the goodwill of the residents of Te\underline{n}pa\d{t}\d{t}i and A\underline{l}ak\={a}puram, the support of faculty and students from the Folklore Department at Madurai Kamaraj University, and the assistance of the Chella Meenakshi Centre for Educational Research and Services. Funding for fieldwork was provided by the US National Science Foundation Doctoral Dissertation Improvement grant (no. BCS-1121326), a Fulbright-Nehru Student Researcher Award, the Stanford Center for South Asia, and a National Science Foundation Interdisciplinary Behavioral \& Social Science Research grant (no. IBSS-1743019). We thank Cristopher Moore and Daniel Larremore for useful discussions and the Santa Fe Institute  for providing the environment fostering these interactions. 
	
\bibliographystyle{apsrev4-1}	
\bibliography{Bibliography}

\newcommand{\beginsupplement}{%
        \setcounter{table}{0}
        \renewcommand{\thetable}{S\arabic{table}}%
        \setcounter{figure}{0}
        \renewcommand{\thefigure}{S\arabic{figure}}%
        \setcounter{equation}{0}
        \renewcommand{\theequation}{S\arabic{equation}}
         \setcounter{section}{0}
        \renewcommand{\thesection}{S\arabic{section}}
 }

\clearpage
\beginsupplement

\section*{{Supporting Information (SI)}}


\section{Methods: EM detailed derivation}
\label{apx:EMderivation}

We show derivations of the updates given in equations ({\color{red} 12})-({\color{red} 15}) of the main manuscript. 

The partial derivative with respect to the elements of the affinity matrices $W^{(\alpha)}$ is given by
\begin{align}
\label{der_w}
\frac{\partial \mathcal{L}}{\partial w_{kl}^{(\alpha)}}= (1-\gamma)\frac{\partial \mathcal{L}_G}{\partial w_{kl}^{(\alpha)}} = (1-\gamma)\sum_{i,j} \bigg[\frac{A_{ij}^{(\alpha)} \rho_{ijkl}^{(\alpha)}}{w_{kl}^{(\alpha)}}-u_{ik}v_{jl}\bigg]\ \ .
\end{align}
The valid update when $\gamma$ is different from 1, is given by setting equation (\ref{der_w}) to zero and we obtain
\begin{align}
\label{eq:w}
w_{kl}^{(\alpha)} = \frac{\sum_{i,j} A_{ij}^{(\alpha)} \rho_{ijkl}^{(\alpha)}}{\sum_i u_{ik} \sum_j v_{jl}} \ \ .
\end{align}

In order to take the derivative with respect to $\beta_{kz}$ we need to consider the Lagrange multiplier $\lambda_k^{(\beta)}$ because of the constraint in equation ({\color{red} 11}). Then,
\begin{align}
\frac{\partial \mathcal{L}}{\partial \beta_{kz}} = \gamma \, \bigg( \frac{1}{\beta_{kz}} \sum_{i} x_{iz} h_{izk}\bigg) - \lambda_k^{(\beta)} \ \ ,
\end{align}
and setting it to zero implies
\begin{align}
\label{eq:b}
\beta_{kz} = \frac{\gamma}{\lambda_k^{(\beta)}} \sum_{i} x_{iz} h_{izk} \ \ .
\end{align}
Enforcing the constraint ({\color{red} 11}), we have 
\begin{align}
\sum_z \frac{\gamma}{\lambda_k^{(\beta)}} \sum_{i} x_{iz} h_{izk}= 1 \ \ ,
\end{align}
which implies
\begin{align}
\label{eq:lambda}
\lambda_k^{(\beta)} = \gamma \sum_{i,z}  x_{iz} h_{izk} \ \ .
\end{align}
Plugging (\ref{eq:lambda}) into (\ref{eq:b}), we obtain the update:
\begin{align}
\label{eq:beta}
    \beta_{kz} = \frac{\sum_{i} x_{iz} h_{izk}}{\sum_{i,z}  x_{iz} h_{izk}} \ \ . 
\end{align}
Focusing the attention on the elements of the matrix $U$, we first consider that plugging the update for $w_{kl}^{(\alpha)}$ given in equation (\ref{eq:w}) into the log-likelihood of the structural dimension $\mathcal{L}_{G}$, the last term becomes a constant. Indeed, 
\begin{align}
\nonumber
- \sum_{i,j}\sum_{k,l}u_{ik}v_{jl}\frac{\sum_{i,j} A_{ij}^{(\alpha)} \rho_{ijkl}^{(\alpha)}}{\sum_i u_{ik} \sum_j v_{jl}} 
&= - \sum_{k,l}\bigg(\frac{\sum_{i,j} A_{ij}^{(\alpha)} \rho_{ijkl}^{(\alpha)}}{\sum_i u_{ik} \sum_j v_{jl}} \sum_{i,j}u_{ik}v_{jl}\bigg)\\
\nonumber
&= - \sum_{k,l}\bigg(\sum_{i,j} A_{ij}^{(\alpha)} \rho_{ijkl}^{(\alpha)}\bigg)\\
\nonumber
&= - \sum_{i,j} A_{ij}^{(\alpha)} \sum_{k,l} \rho_{ijkl}^{(\alpha)}\\
\nonumber
&= - \sum_{i,j} A_{ij}^{(\alpha)} \\
&= - T^{(\alpha)}
\end{align}
since $\sum_{k,l} \rho_{ijkl}^{(\alpha)}=1$ and $\sum_{i,j} A_{ij}^{(\alpha)}$ is the number of links in layer $\alpha$ when the network is directed (or twice this value in the undirected case). Thus, we can ignore this term when estimating $u_{ik}$. 
Using the same strategy used in computing the update of $\beta$, we compute the Lagrange multiplier $\lambda_i^{(u)}$ for the constraint in equation  ({\color{red} 11}). Then,
\begin{align}
\frac{\partial \mathcal{L}}{\partial u_{ik}} = \frac{1}{u_{ik}}\bigg(\gamma \sum_z x_{iz} h_{izk} + (1-\gamma) \sum_{j,l,\alpha}A_{ij}^{(\alpha)} \rho_{ijkl}^{(\alpha)}\bigg) - \lambda_i^{(u)}
\end{align}
and 
\begin{align}
    \label{eq:u}
u_{ik} = \frac{1}{\lambda_i^{(u)}}\bigg(\gamma \sum_z x_{iz} h_{izk} + (1-\gamma) \sum_{j,l,\alpha}A_{ij}^{(\alpha)} \rho_{ijkl}^{(\alpha)}\bigg) \ \ .
\end{align}
Enforcing the constraint $\sum_{k} u_{ik}=1$ yields:
\begin{align}
\sum_k \frac{1}{\lambda_i^{(u)}}\bigg(\gamma \sum_z x_{iz} h_{izk} + (1-\gamma) \sum_{j,l,\alpha}A_{ij}^{(\alpha)} \rho_{ijkl}^{(\alpha)}\bigg) = 1 
\end{align}
which implies
\begin{align}
\nonumber
\lambda_i^{(u)} &= \sum_k\bigg(\gamma \sum_z x_{iz} h_{izk} + (1-\gamma) \sum_{j,l,\alpha}A_{ij}^{(\alpha)} \rho_{ijkl}^{(\alpha)}\bigg)\\
\label{eq:delta}
&= \gamma + (1-\gamma) \sum_{j,\alpha}A_{ij}^{(\alpha)} \ \ ,
\end{align}
since $\sum_{k} h_{izk}=1$,  $\sum_{k,l} \rho_{ijkl}^{(\alpha)}=1$ and $\sum_{z} x_{iz}=1$.
Plugging the result of equation (\ref{eq:delta}) into the equality (\ref{eq:u}) we obtain
\begin{align}
\label{eq:u2}
    u_{ik} = \frac{\gamma \sum_z x_{iz} h_{izk} + (1-\gamma) \sum_{j,l,\alpha}A_{ij}^{(\alpha)} \rho_{ijkl}^{(\alpha)}}{ \gamma + (1-\gamma) \sum_{j,\alpha}A_{ij}^{(\alpha)}} \ \ . 
\end{align}
In order to compute the update for $V$, we fix $j$ and $l$, rewriting the attribute dimension $\mathcal{L}_{X}(U,V,\beta,h)$  as
\begin{align}
\mathcal{L}_{X}(U,V,\beta,h)=\sum_{j,z,l} x_{jz} (h_{jzl}\,\text{log}(\beta_{lz} (u_{jl}+v_{jl})) - h_{jzl}\,\text{log}(h_{jzl}) ) \ \ .
\end{align}
Analogously to before, we consider the Lagrange multiplier to satisfy the constraint given in equation ({\color{red} 11}), and we obtain
\begin{align}
\label{eq:v}
	    v_{jl} = \frac{\gamma \sum_z x_{jz} h_{jzl} + (1-\gamma) \sum_{i,k,\alpha}A_{ij}^{(\alpha)} \rho_{ijkl}^{(\alpha)}}{ \gamma + (1-\gamma) \sum_{i,\alpha}A_{ij}^{(\alpha)}} \ \ . 
\end{align}
In each iteration of the EM algorithm, the parameters in $\Theta = ({U,V, W,\beta})$ are updated with  equations (\ref{eq:w}), (\ref{eq:beta}), (\ref{eq:u2}) and (\ref{eq:v}), until the log-likelihood $\mathcal{L}(\Theta, h,\rho)$ reaches a fixed point. 


\section{Computational complexity}
\label{apx:ComputationalComplexity}
\begin{figure}[htp]%
\centering
\centering
    \includegraphics[width=\linewidth]{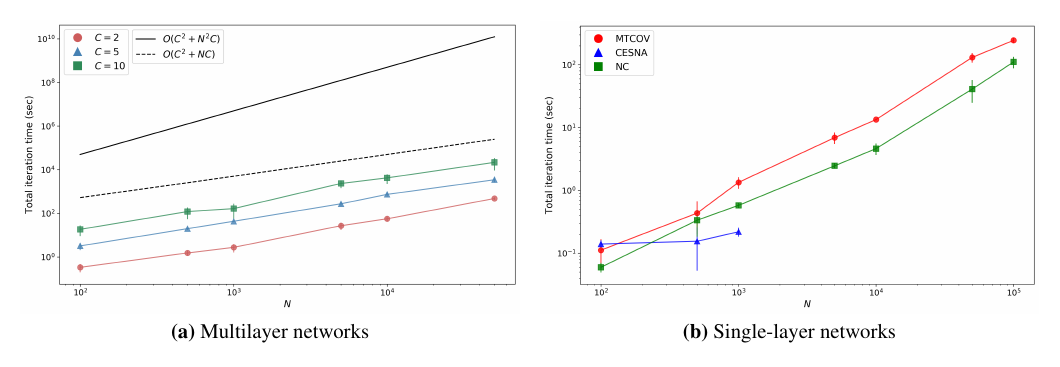}
 \caption{Computational complexity. (a) Total iteration time of \mtcov\text{} on synthetic multilayer networks. (b) Total iteration time of \mtcov, \cesna\text{} and \nc\text{} on synthetic single-layer networks. Results are averages and standard deviations over 5 different random initial conditions.     }
 \label{fig:times}%
\end{figure}

Based on Algorithm 1 in the main manuscript, the computational complexity per iteration of MTCOV scales as $O(MC^2+NCZ)$, where $M$ is the number of edges summed across layers, $C$ the number of communities, $N$ the number of nodes and $Z$ the number of categories of the categorical attribute. In practice, $C$ and $Z$ have similar order of magnitude, usually much smaller than the system size $M$; for sparse networks, as is often the case for real datasets, $M \propto N$, thus the algorithm is highly scalable with a total running time linear in the system size. Figure \ref{fig:times} shows the empirical study of the computational complexity on large scale synthetic networks, both single and multilayer.\\
We generate synthetic multilayer networks as described in the Results section of the main manuscript, in the subsection Multilayer synthetic networks with ground truth. We generate directed networks with $L=4$ layers, one assortative, one disassortative, one core-periphery and one with biased directed structure. The number of categories of the categorical attribute is $Z=2$ and the match between attributes and planted communities is equal to $0.5$. We vary the number of nodes $N \in \{100, 500, 1000, 5000, 10000, 50000\}$ and the number of communities $C \in \{2, 5, 10\}$. We run \mtcov\text{} with 5 different random initializations and the average results are shown in Fig. \ref{fig:times} (a). The black lines are the baseline for the quadratic (solid line) and the linear (dashed line) complexity in $N$, with $C=5$ (similar results are obtained with different values of $C$). As expected, the total iteration time of \mtcov\text{} is linear in the system size.\\
For single-layer networks, we generate synthetic data by using the approach described in the subsection \ref{apx: synthetic_single}. We generate undirected networks with $C=2$ communities, $Z=2$ categories and a match between attributes and planted communities equal to $0.5$. The $p_{in}=\frac{16}{N}$, $p_{out}=\frac{4}{N}$ and the number of nodes varies $N \in \{100, 500, 1000, 5000, 10000, 50000, 100000\}$. Figure \ref{fig:times} (b) shows the performance of \mtcov\text{} in comparison to \cesna\text{} and \nc\text{}; the results are averaged over $5$ random initializations. \cesna\text{} has convergence issues for networks with $N>1000$, which results in unreliable community detection on large networks, hence we omit those results. \nc\text{} and \mtcov\text{} show a linear computational complexity in the system size.

\section{Multilayer social support network of rural Indian villages}
\label{apx: multi_real_data}

\subsection{Normalization}
\label{apx: normalization}
A way to control the magnitude of the likelihood terms is rescaling each term individually. Here, we estimate two linear regression models in order to obtain the normalization constants for the two terms. Given the access to several network datasets of the same kind (social support networks in Indian villages), we collect log-likelihood values with respect to the number of nodes ($N$), edges ($E$) of the observed network and the number of categories of the categorical attribute ($Z$). Quantitatively, this means normalizing as:
\begin{align}
\label{normalization_eq}
    \mathcal{L} = (1-\gamma)\frac{\mathcal{L}_G}{c^G_N N+c^G_E E+c^G_Z Z} + \gamma \frac{\mathcal{L}_X}{c^X_N N+c^X_E E+c^X_Z Z} \ \ .
\end{align}
The super and the subscripts of the $c$ parameters indicate the dependent variable ($\mathcal{L}_X$ or $ \mathcal{L}_G$) and the input regressor they refer to, respectively. In particular, we collect the data by running the model for each pair of network and categorical attribute, arbitrarily fixing the number of communities $C=3$ and the scaling parameter $\gamma=0.5$. 
Table \ref{tab: Estimation} states the values of the statistically significant coefficients for the estimation of the log-likelihood terms, and only those coefficients are used in the normalized equation (\ref{normalization_eq}).
\begin{table}[ht]
\centering
\resizebox{0.3\textwidth}{!}{
\renewcommand\arraystretch{1.25}
\begin{tabular}{ccc}
\clineB{1-3}{3}
\textbf{} & \textbf{$\mathcal{L}_X$} & \textbf{$\mathcal{L}_G$} \\ \hline
$c_N$         & $-0.486^{***} $     & $-1.778^{**} $     \\
$c_E$         &             & $-6.158^{***}  $    \\
$c_Z$         & $-33.862^{***}$    &             \\ 
\clineB{1-3}{3}
\end{tabular}}
\caption{Coefficient estimates $c^X_N$, $c^X_E$, $c^X_Z$, $c^G_N$, $c^G_E$ and $c^G_Z$ for the two linear regression models.}
\label{tab: Estimation}
\end{table}
 
On one side, this procedure allows to obtain coefficient estimates useful for analyzing all four networks of Indian villages in a quantitative and automatized way. On the other side, we are aware that these coefficients are strictly related to the type of network we are working with, i.e. their values cannot be used in network datasets other than the social support networks used here. Future works should investigate an automated normalization procedure applicable to any network dataset as a pre-processing step.

\subsection{Hyperparameter fitting}
\label{apx: hyperparameters}
\begin{table}[h!]
\centering
\resizebox{0.8\textwidth}{!}{
\renewcommand\arraystretch{1.25}
\begin{tabular}{ll|cccc}
\clineB{1-6}{3}
\multicolumn{1}{c}{\textbf{}}  &\multicolumn{1}{c}{\textbf{}}  &\multicolumn{4}{c}{\textbf{Hyperparameters setting}}    \\
\multicolumn{1}{l}{\textbf{Attribute}} &\multicolumn{1}{l|}{\textbf{Method}} & \textbf{Ala 2013}           & \textbf{Ala 2017}       & \textbf{Ten 2013} & \textbf{Ten 2017}   \\ \hline
\multicolumn{1}{l}{\textbf{}}&\multicolumn{1}{l|}{\mt}& $C=8$ & $C=9$ & $C=3$& $C=3$\\ 
\multirow{1}{*}{Caste} &\multicolumn{1}{l|}{\mtcov } & $C=7$, $\gamma=0.8$ & $C=7$, $\gamma=0.8$ & $C=6$, $\gamma=0.8$&$C=6$, $\gamma=0.9$\\ 
\multirow{1}{*}{Religion}  &\multicolumn{1}{l|}{\mtcov } &$C=6$, $\gamma=0.8$   &$C=6$, $\gamma=0.8$   &  $C=6$, $\gamma=0.7$ & $C=6$, $\gamma=0.7$ \\  
\multirow{1}{*}{Age}  &\multicolumn{1}{l|}{\mtcov } & $C=8$, $\gamma=0.4$ & $C=8$, $\gamma=0.3$ & $C=9$, $\gamma=0.2$ & $C=7$, $\gamma=0.3$\\  
\multirow{1}{*}{Gender}  &\multicolumn{1}{l|}{\mtcov } & $C=10$, $\gamma=0.8$ & $C=10$, $\gamma=0.7$ & $C=10$, $\gamma=0.4$ & $C=9$, $\gamma=0.7$\\  
\clineB{1-6}{3} 
\end{tabular}}
\caption{Values of the hyperparameters $C$ and $\gamma$ extracted by 5-fold cross-validation combined with grid-search.}
\label{tab: estimates}
\end{table}

\subsection{Biased link prediction}
\label{apx: biased_tpe}
We use sampling bias techniques to assign higher or lower sampling probability to the entries of the adjacency tensor, which correspond to edges and non-edges. By defining \textit{tpe} the total probability of selecting one edge (non-zero entry), we assign to the entries $a_{ij}^{(\alpha)}>0$ the probability of being selected in the test set given by: 
\begin{align}
    p_1 = \frac{\textit{tpe}}{E} \ \ ,
\end{align}
and for $0$ entries
\begin{align}
    p_2 =\frac{1- \textit{tpe}}{N^{2}\, L - E} \ \ ,
\end{align}
where $E$ and $N^2\, L$ are the total number of the edges and entries of the adjacency tensor respectively. These two probabilities are assigned to the entries $a_{ij}^{(\alpha)}$ to perform a biased selection while choosing test and train sets, as in a selection of a binary mask. The \textit{tpe} is used to select entries for the test set; in case of selecting entries uniformly at random, this value would be around $0.004$. This low value is due to the common case in real networks of having sparse matrices, where the number of non-zero entries is much lower than the number of zeros. We create three different situations, starting from $tpe =0.001$ where the probability of selecting one edge is lower than the probability of choosing one non edge, and the number of edges in the training set is much higher than the number in the test set. Then we have $tpe =0.015$ and finally $tpe =0.03$, where the probability of selecting one edge in the test set is higher than the probability of choosing one non edge, and the test set has a bigger number of positive examples. These settings follow an increasing order of complexity, starting from an under-represented case, where $tpe=0.001$, and ending with a difficult task where the number of edges in the test set is over-represented, $tpe=0.03$. We run 10 independent trials for each setting and model.


\begin{table}[ph!]
\centering
\resizebox{0.9\textwidth}{!}{
\renewcommand\arraystretch{1.25}
\begin{tabular}{cll|cccc}
\clineB{1-7}{3}
\multicolumn{1}{c}{\textbf{}}  &\multicolumn{1}{c}{\textbf{}}  &\multicolumn{1}{c}{\textbf{}}        &\multicolumn{4}{c}{\textbf{AUC for link prediction}}    \\
\multicolumn{1}{l}{\textbf{Edge bias sampling}} &\multicolumn{1}{l}{\textbf{Attribute}} &\multicolumn{1}{l|}{\textbf{Method}} & \textbf{Ala 2013}           & \textbf{Ala 2017}           & \textbf{Ten 2013} & \textbf{Ten 2017}   \\ \hline
\multirow{4}{*}{$tpe=0.001$} & &\multicolumn{1}{l|}{\mt}& 0.82$\pm$0.01 & 0.85$\pm$0.01 &0.78$\pm$0.02&0.82$\pm$0.02\\ 
&\multirow{1}{*}{Caste}  &\multicolumn{1}{l|}{\mtcov } & \textbf{0.851$\pm$0.008} & \textbf{0.87$\pm$0.01}&\textbf{0.85$\pm$0.01}&\textbf{0.83$\pm$0.02}\\ 
&\multirow{1}{*}{Religion}  &\multicolumn{1}{l|}{\mtcov } & 0.82$\pm$0.02 &  0.85$\pm$0.02 & 0.83$\pm$0.02 & 0.83$\pm$0.01\\  
&\multirow{1}{*}{Age}  &\multicolumn{1}{l|}{\mtcov } & 0.83$\pm$0.01 & 0.83$\pm$0.01& 0.81$\pm$0.02 &0.821$\pm$0.009\\  
&\multirow{1}{*}{Gender}  &\multicolumn{1}{l|}{\mtcov } &0.81$\pm$0.02 &0.85$\pm$0.02&0.83$\pm$0.01&0.82$\pm$0.01\\  
\clineB{1-7}{3}
\multirow{4}{*}{$tpe=0.004$} & &\multicolumn{1}{l|}{\mt}& 0.771$\pm$0.009 & 0.835$\pm$0.006 &0.758$\pm$0.005&0.81$\pm$0.01\\
&\multirow{1}{*}{Caste}  &\multicolumn{1}{l|}{\mtcov } & \textbf{0.846$\pm$0.007} & \textbf{0.865$\pm$0.006} &\textbf{0.838$\pm$0.006}&\textbf{0.82$\pm$0.01}\\ 
&\multirow{1}{*}{Religion}  &\multicolumn{1}{l|}{\mtcov } & 0.816$\pm$0.009 & 0.83$\pm$0.01 &0.81$\pm$0.01& 0.82$\pm$0.01\\  
&\multirow{1}{*}{Age}  &\multicolumn{1}{l|}{\mtcov } & 0.82$\pm$0.01 & 0.83$\pm$0.01 & 0.791$\pm$0.009 &0.79$\pm$0.02\\  
&\multirow{1}{*}{Gender}  &\multicolumn{1}{l|}{\mtcov } & 0.790$\pm$0.007& 0.83$\pm$0.02& 0.81$\pm$0.01&0.82$\pm$0.02\\  
\clineB{1-7}{3}
\multirow{4}{*}{$tpe=0.015$} & &\multicolumn{1}{l|}{\mt} & 0.62$\pm$0.01 & 0.73$\pm$0.01 &0.68$\pm$0.01&0.76$\pm$0.01\\ 
&\multirow{1}{*}{Caste}  &\multicolumn{1}{l|}{\mtcov } & \textbf{0.806$\pm$0.008} &\textbf{0.847$\pm$0.006} &\textbf{0.811$\pm$0.007}&\textbf{0.812$\pm$0.006}\\ 
&\multirow{1}{*}{Religion}  &\multicolumn{1}{l|}{\mtcov } & 0.76$\pm$0.01 &0.81$\pm$0.01  &0.75$\pm$0.01&0.78$\pm$0.01 \\ 
&\multirow{1}{*}{Age}  &\multicolumn{1}{l|}{\mtcov } & 0.68$\pm$0.01 & 0.77$\pm$0.02 & 0.68$\pm$0.02 &0.74$\pm$0.03\\  
&\multirow{1}{*}{Gender}  &\multicolumn{1}{l|}{\mtcov } &0.71$\pm$0.01 &0.79$\pm$0.02&0.70$\pm$0.01& 0.76$\pm$0.02\\
\clineB{1-7}{3}
\multirow{4}{*}{$tpe=0.03$} & &\multicolumn{1}{l|}{\mt}& 0.46$\pm$0.01 & 0.554$\pm$0.009 &0.55$\pm$0.01&0.63$\pm$0.01\\ 
&\multirow{1}{*}{Caste}  &\multicolumn{1}{l|}{\mtcov } & \textbf{0.73$\pm$0.01} &\textbf{0.79$\pm$0.01}&\textbf{0.760$\pm$0.008}&\textbf{0.77$\pm$0.01}\\ 
&\multirow{1}{*}{Religion}  &\multicolumn{1}{l|}{\mtcov } & 0.68$\pm$0.02 &0.74$\pm$0.02 &0.64$\pm$0.02 &0.68$\pm$0.01\\ 
&\multirow{1}{*}{Age}  &\multicolumn{1}{l|}{\mtcov } & 0.54$\pm$0.02 & 0.63$\pm$0.02 & 0.54$\pm$0.01 &0.61$\pm$0.02\\  
&\multirow{1}{*}{Gender}  &\multicolumn{1}{l|}{\mtcov } &0.58$\pm$0.01 &0.66$\pm$0.02&0.55$\pm$0.02& 0.65$\pm$0.01\\  
\clineB{1-7}{3}
\end{tabular}}
\caption{Prediction performance on real multilayer networks with attributes and bias sampling on edges. Results are averages and standard deviations over 10 independent trials of cross-validation with 80-20 splits applied to the entries of $A$ (the whole $X$ is given in input); best performances are in boldface. Datasets are described in the main manuscript.}
\label{tab: bias}
\end{table}

\section{Single-layer networks experiments}
\label{apx: single_data}
We generate 10 independent realization with different random initializaton for each network. We run \nc\text{} setting the maximum number of BP steps before aborting to 40 and maximum number of EM steps before aborting to 500 to be consistent with \mtcov. \mtcov\text{} and \cesna\text{} use the fraction match as weight between the network and attributes. Moreover, for \cesna\text{} we add a new community with all non-classified nodes, and this is done for all tests. Therefore, \nc\text{} and \mtcov\text{} give the possibility to decide if considering a hard membership or a mixed one, thanks to the posterior probabilities given as output, and assigning the community with the highest probability, even though \nc\text{} doesn't mention the possibility of mixed-membership in the paper. On the other hand, \cesna\text{} does not return those values, and we have to work with their output files which provide an overlapping partition of the nodes.

\subsection{Single-layer synthetic networks}
\label{apx: synthetic_single}
\begin{figure}[]
  \centering
    \includegraphics[width=\linewidth]{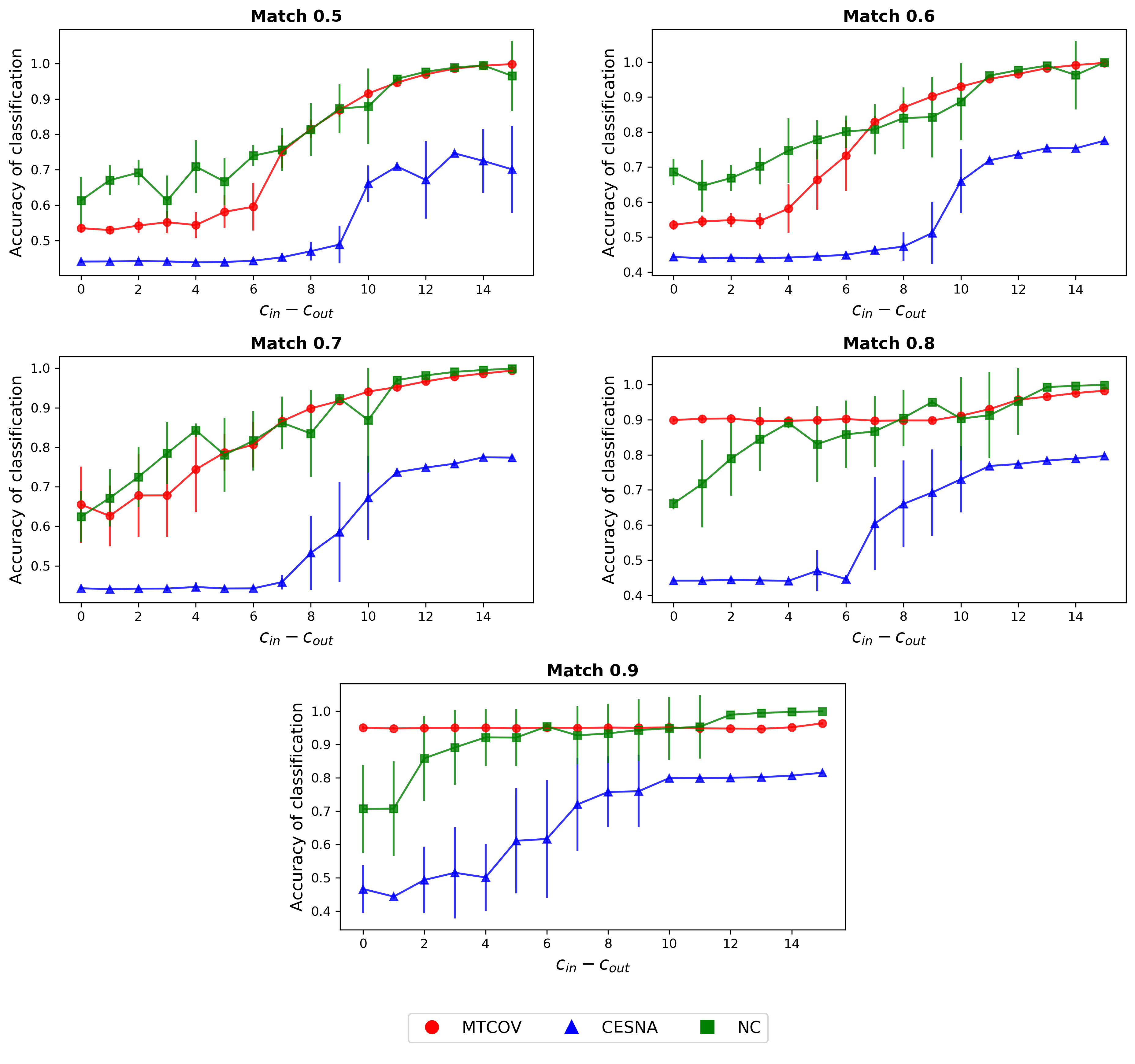}
\caption{Accuracy of classification for synthetic single-layer networks with two communities of equal size, generated with the stochastic block model. Each plot shows results with a given match between metadata and planted communities. The results are averaged over 10 independent trials and the bars represent the standard deviation. The accuracy is measured with F1-score as similarity measure. }
\label{fig:synthetic}
\end{figure}

In analogy with what done for multilayer synthetic networks, we test \mtcov 's ability to detect communities on synthetic single-layer networks, using the approach proposed by Newman and Clauset \cite{newman2016structure}, where they generate synthetic networks with known community structure embedded within them. Networks are generated using the stochastic block model \cite{holland1983stochastic, karrer2011stochastic},  with $N=1000$ nodes and $C=2$ non-overlapping communities of equal size. Edge probabilities are $p_{in} = c_{in}/n$ and $p_{out} = c_{out}/n$, for within-group and between-group edges, respectively. The difference $c_{in}-c_{out}$ measures the strength of the community structure, when $c_{in}$ is much greater than $c_{out}$ the communities are easy to detect from network structure alone, and it becomes harder when these two quantities are close. Discrete-valued attributes are generated on nodes, which match the true community assignments of nodes a given fraction of the times, and are instead chosen uniformly at random from the non-matching values otherwise. This allows to control the correlation between attributes and community structure and hence test the algorithm's ability to exploit the extra information of varying quality. The match between attributes and planted communities varies between $0.5$ and $0.9$, the higher this value the higher the extent to which node attributes help predicting edges.
In Fig. \ref{fig:synthetic} we show the fraction of correctly classified nodes in terms of F1-score (results in terms of Jaccard are similar) for such experiments. We notice first a clear pattern where all the methods increase their performance and reduce their variance as the difference  $c_{in}-c_{out}$ gets bigger, going from a hard to an easy to detect regime. In the hard regime, where community structure weakens,  both \mtcov\text{}  and \nc\text{} remain robust in detecting the communities for scenarios where attributes are correlated. However, \mtcov\text{}  shows lower variance and has more stable results for high attribute correlations. In addition, empirically we observe that \nc\text{} does not reach converge in $19\%$ of the trials, while \mtcov\text{}  only for the $13\%$ of the times. \cesna\text{} always shows lower performances, probably penalized by the relative high number of non classified examples which we also observe experimentally. 

\subsection{Single-layer real networks}
\paragraph{The \textit{facebook} ego-networks} We consider the attributes \textit{Birthday (B), Hometown (H)} and \textit{Location (L)}, relationship that potentially correlate with the community partitions. In addition, they have a reasonable number of values, compared to other attributes whose proportions of missing values are too high. We take in consideration also the sum by row: since we are assuming a multinomial distribution for the attribute, we ask for a sum by row equal to 1. As first step, we combine all the 10 ego networks merging both the edges and the covariates. In this way we double check the real number of edges and the number of nodes. Moreover, we build complete design matrices for B, H and L. They are used for retrieving the largest possible number of attributes for each node. However, we decide to work with the ego networks separately, for having a clear classification of the ground truth. We build the input files using the following procedure, where we consider only the nodes having at least one edge and the attribute. We obtain 30 networks: 3 for each ego network according to the attribute B, H or L; and for each combination of ego and attribute we have different adjacency matrix $A$ and ground truth $circles$.  However, we analyse 21 networks because we have to discard three ego networks (3437, 3980, 698) due to the null number of communities or to the small number of nodes with ground truth.
 
 \paragraph{American College Football}  We use the corrected version of the dataset \cite{football_corrected}, where among all they assign the independent teams to a unique community label rather than assigning them a single community label as in the original football dataset. In this way the number of communities given by the number of conferences is 19.
 
 \paragraph{Political Blogs (polblogs)} We remove the isolated vertices and self-loops. The ground-truth communities are \textit{left/liberal, right/conservative}, so $C=2$. 

For the analysis of these networks, we run \cesna\text{} with the default parameter $\alpha=0.5$ because their released code does not allow to perform a cross-validation procedure on the scaling parameter. The facebook networks have overlapping communities, while for the other two datasets we assume non-overlapping,  according to the proposed ground-truth. For facebook we consider the 21 ego networks as different iterations.  For the other two, since both \mtcov\text{}  and \nc\text{} are based on a EM algorithm which does not ensure to reach a global maximum, we perform 10 restarts of the algorithms with different initializations at random. Results are presented in the main manuscript.


\end{document}